\documentclass[aps,apl,onecolumn,preprint,superscriptaddress,showkeys,nofootinbib,twoside,a4paper,amsfont,amssymb,amsmath]{revtex4-1}
\usepackage{color}
\usepackage{graphicx}% Include figure files
\usepackage{dcolumn}% Align table columns on decimal point
\usepackage{bm}% bold math
\usepackage{comment}
\usepackage[english]{babel}
\usepackage[latin1]{inputenc}
\usepackage[bottom=2.5cm,top=2.5cm, left=2cm, right=2cm]{geometry}
\usepackage{siunitx}
\newcommand{\SIadj}[2]{\SI[number-unit-product={\text{-}}]{#1}{#2}}
\DeclareSIUnit\angstrom{\text {\AA}}
\DeclareSIUnit\bar{\text bar}
\DeclareSIUnit\eV{\text eV}
\usepackage{xr}

\usepackage{pdfpages}

\makeatletter
\AtBeginDocument{\let\LS@rot\@undefined}
\makeatother

\makeatletter
\newcommand*{\addFileDependency}[1]{
  \typeout{(#1)}
  \@addtofilelist{#1}
  \IfFileExists{#1}{}{\typeout{No file #1.}}
}
\makeatother

\newcommand*{\myexternaldocument}[1]{
    \externaldocument{#1}
    \addFileDependency{#1.tex}
    \addFileDependency{#1.aux}
}
\myexternaldocument{Supplementary_Materials_Article_class}

\usepackage[unicode=true, bookmarks=true, bookmarksnumbered=false, bookmarksopen=false, breaklinks=false, pdfborder={0 0 1}, backref=false, colorlinks=false]{hyperref}
\hypersetup{pdftitle={Domain-wall-to-domain interconversion in a polar conducting oxide}, pdfauthor={{M. F. Sarott, M. Ahmadi, J. van Gent Gonz\'{a}lez, E. van der Veer, M. C. Weber, J. Pinna, D. A. Chaney, A. Sinha, H. R. J. Cox, T. Perrault A. Sabet Ghorabaei, M. A. Loi, P. Ghosez, B. J. Kooi, B. Noheda}, pdfkeywords={Keywords}}}

\begin{document}

\title{Domain-wall-to-domain interconversion in a polar conducting oxide}

\author{Martin F.\ Sarott}
\email{m.f.sarott@rug.nl}
\affiliation{Zernike Institute for Advanced Materials, University of Groningen, 9747AG Groningen, The Netherlands.}
\affiliation{Groningen Cognitive Systems and Materials Center (CogniGron), University of Groningen, 9747AG Groningen, The Netherlands.}

\author{Majid Ahmadi}
\affiliation{Zernike Institute for Advanced Materials, University of Groningen, 9747AG Groningen, The Netherlands.}
\affiliation{Groningen Cognitive Systems and Materials Center (CogniGron), University of Groningen, 9747AG Groningen, The Netherlands.}

\author{Johanna van Gent Gonz\'{a}lez}
\affiliation{Zernike Institute for Advanced Materials, University of Groningen, 9747AG Groningen, The Netherlands.}

\author{Ewout van der Veer}
\affiliation{Zernike Institute for Advanced Materials, University of Groningen, 9747AG Groningen, The Netherlands.}
\affiliation{Faculty of Physics, University of Duisburg-Essen, Duisburg, Germany.}
\affiliation{Research Center Future Energy Materials and Systems (RC FEMS), Research Alliance Ruhr, Bochum, Germany.}

\author{Mads C. Weber}
\affiliation{Institut des Mol\'{e}cules et Mat\'{e}riaux du Mans, UMR 6283 CNRS, Le Mans Universit\'{e}, 72085 Le Mans, France.}

\author{Jacopo Pinna}
\affiliation{Zernike Institute for Advanced Materials, University of Groningen, 9747AG Groningen, The Netherlands.}

\author{Daniel A. Chaney}
\affiliation{The European Synchrotron Radiation Facility (ESRF), Grenoble, France.}

\author{Arijit Sinha}
\affiliation{Theoretical Materials Physics, Q-MAT, University of Li\`{e}ge, Li\`{e}ge, Belgium.}

\author{Hamideh Hassani}
\affiliation{Theoretical Materials Physics, Q-MAT, University of Li\`{e}ge, Li\`{e}ge, Belgium.}

\author{Horatio R. J. Cox}
\affiliation{Zernike Institute for Advanced Materials, University of Groningen, 9747AG Groningen, The Netherlands.}
\affiliation{Groningen Cognitive Systems and Materials Center (CogniGron), University of Groningen, 9747AG Groningen, The Netherlands.}

\author{Thomas Perrault}
\affiliation{Institut des Mol\'{e}cules et Mat\'{e}riaux du Mans, UMR 6283 CNRS, Le Mans Universit\'{e}, 72085 Le Mans, France.}

\author{Amir Sabet Ghorabaei}
\affiliation{Zernike Institute for Advanced Materials, University of Groningen, 9747AG Groningen, The Netherlands.}

\author{Maria A. Loi}
\affiliation{Zernike Institute for Advanced Materials, University of Groningen, 9747AG Groningen, The Netherlands.}
\affiliation{Groningen Cognitive Systems and Materials Center (CogniGron), University of Groningen, 9747AG Groningen, The Netherlands.}

\author{Philippe Ghosez}
\affiliation{Theoretical Materials Physics, Q-MAT, University of Li\`{e}ge, Li\`{e}ge, Belgium.}

\author{Bart J.\ Kooi}
\affiliation{Zernike Institute for Advanced Materials, University of Groningen, 9747AG Groningen, The Netherlands.}
\affiliation{Groningen Cognitive Systems and Materials Center (CogniGron), University of Groningen, 9747AG Groningen, The Netherlands.}

\author{Beatriz Noheda}
\email{b.noheda@rug.nl}
\affiliation{Zernike Institute for Advanced Materials, University of Groningen, 9747AG Groningen, The Netherlands.}
\affiliation{Groningen Cognitive Systems and Materials Center (CogniGron), University of Groningen, 9747AG Groningen, The Netherlands.}

\date{October 8, 2026}

\begin{abstract}

Domain walls in ferroic crystals are topological defects with a symmetry distinct from the domain interior, allowing them to display functionalities that are otherwise forbidden in the host material. 
Despite their unique properties, domain walls are strictly confined to two-dimensional interfaces, which hinders unveiling their true nature and limits their applicability. 
Here, we demonstrate reversible dimensionality control by expanding nanoscale domain walls into three-dimensional functional regions in polar epitaxial tungsten trioxide ($\text{WO}_3$).
This domain-wall-to-domain interconversion is enabled by the conducting, non-polar domain walls adopting the parent-phase structure. 
An alternating electric field applied through a scanning probe tip induces localized heating, which expands this domain-wall phase into the adjacent polar domains. 
This approach enables reversible spatial control to switch between mutually exclusive properties of domains (polarity) and domain walls (conductivity), offering a new paradigm for reconfigurable oxide electronics.

\end{abstract}

%\keywords{Keywords}
\maketitle
%%%%%%%%%%%%%%%%%%%%%%%%%%%%%%%%%%%%%%%%%%%%%%%%%%%%%%%%%%%%%%%%%%%%%%%%%%%%%%%%%%%%%%%%%%%
%Introduction
%%%%%%%%%%%%%%%%%%%%%%%%%%%%%%%%%%%%%%%%%%%%%%%%%%%%%%%%%%%%%%%%%%%%%%%%%%%%%%%%%%%%%%%%%%%
Traditionally, the functionality of ferroic materials was ascribed to the spatial arrangement of three-dimensional domains, i.e.\ regions where the order parameter (e.g.\ polarization, magnetization, or strain) is uniform, and their response to external stimuli.
Over the past two decades, this paradigm has shifted to include the spatially confined domain walls, in which the ferroic order parameter reorients.~\cite{Seidel2009,Catalan2012,Meier2021}
As two-dimensional topological defects with a symmetry different from the bulk, domain walls can be the source of physical phenomena different from the host matrix, including emergent polarization, magnetization, free-electron gases, (super-)conductivity, or magnetoelectric coupling.~\cite{Aird1998,Fiebig2002,Seidel2009,VanAert2011,Guyonnet2011,Sluka2013,Farokhipoor2014,Tselev2016,Frenkel2017,Juraschek2019,Giraldo2021}
These localized emergent properties, along with the ability to create, manipulate, and erase domain walls via external fields, have driven intense efforts to develop nanoelectronic devices that utilize domain walls as active functional elements.~\cite{Catalan2012}\\

So far, the route to enhance the relevance of domain-wall properties in the overall material response was to control the domain-wall density.
This, however, still results in a mixed-phase scenario, where the macroscopic responses are strongly influenced by the properties of the host matrix.
Controlling the spatial dimensions of these localized properties by reversibly expanding domain walls into three-dimensional domains and contracting them back into a confined state would remove this limitation and expand the scope of domain-wall electronics.
Successfully navigating this spatial interconversion could, for instance, unlock intriguing multifunctionalities by switching between mutually exclusive properties, such as conductivity and polarity.~\cite{Anderson1965,Benedek2016}
Conversely, converting the state of a bulk domain into a confined domain wall could enable memory devices with an increased storage density.~\cite{Salje2010}\\

% Theoretically, a gradual transformation between ferroic domain and domain-wall phases can occur across a first-order phase transition if the domain wall adopts a stable structure of the material accessible through a phase transition.~\cite{Mitsek1970,Belov1976,Lajzerowicz1981,Kharchenko1989}
% Experimentally, however, such a continuous interconversion has only been demonstrated in a magnetic material at cryogenic temperatures~\cite{Gnatchenko1981,Hassanpour2021} and once in a structural system at the anti-phase boundaries of a hexagonal tungsten bronze.~\cite{Tao1988}
% Further experimental validation remains elusive, but recent simulations on $\text{CaTiO}_3$ have predicted a broadening of polar domain walls under an applied electric field.~\cite{Lu2025}\\

In this study, we report the interconversion between domains and domain walls in a polar ferroelastic material close to a structural phase transition at room temperature. 
Specifically, we demonstrate the reversible expansion of highly confined, conducting domain walls into extended conducting domains in epitaxial thin films of tungsten trioxide ($\text{WO}_3$).
In their pristine state, these $\text{WO}_3$ films feature a polar triclinic structure with in-plane polarized domains.~\cite{vanderVeer2026}
By applying an AC voltage to the conducting domain walls using a scanning probe tip, we trigger a localized phase transition between the non-conducting polar domain phase and a highly-conducting non-polar domain-wall phase.
Finally, we exploit this interconvertibility between the mutually exclusive properties polarity and conductivity, and highlight its potential for synaptic devices with tunable electrical properties.
While theoretically a gradual transformation between ferroic domain and domain-wall phases has been predicted to occur across a first-order phase transition~\cite{Mitsek1970,Belov1976,Lajzerowicz1981,Kharchenko1989}, experimentally, it has only been demonstrated in a magnetic material at around \SI{2}{\kelvin}.~\cite{Gnatchenko1981,Hassanpour2021}
Our work, thus, constitutes the first realization of such a continuous interconversion in a structural system at room temperature.
%%%%%%%%%%%%%%%%%%%%%%%%%%%%%%%%%%%%%%%%%%%%%%%%%%%%%%%%%%%%%%%%%%%%%%%%%%%%%%%%%%%%%%%%%%%
%End of introduction, Begin of Figure 1
%%%%%%%%%%%%%%%%%%%%%%%%%%%%%%%%%%%%%%%%%%%%%%%%%%%%%%%%%%%%%%%%%%%%%%%%%%%%%%%%%%%%%%%%%%%
\section*{Reversible domain-wall expansion in $\text{WO}_3$}
\begin{figure}[ht!]
\includegraphics[width=\columnwidth]{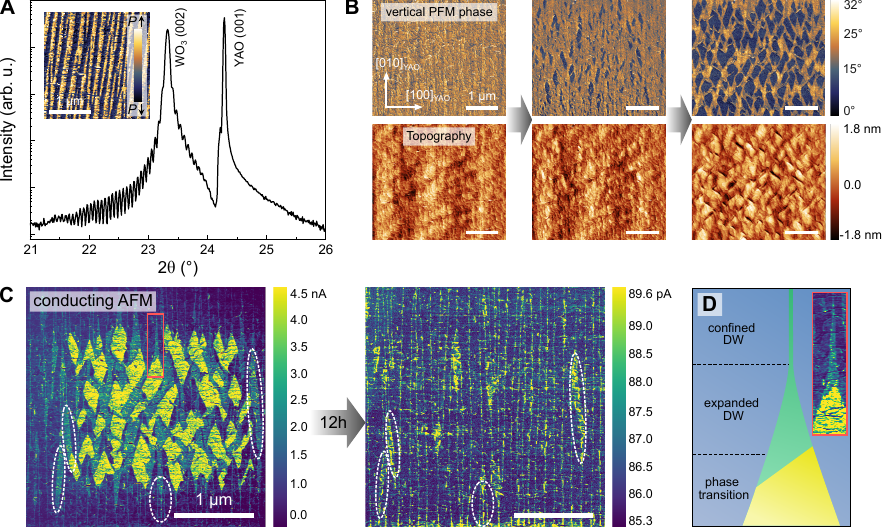}
\caption{\textbf{Expansion of conducting domain walls in polar WO\textsubscript{3} thin films.}
(\textbf{A}) Symmetric X-ray diffraction $\theta$-$2\theta$ measurement of the epitaxial $\text{WO}_3$ film around the YAO 001 peak, displaying pronounced Laue oscillations. 
The lateral PFM phase image in the inset shows the polar \SI{180}{\degree} stripe domains of the pristine film.~\cite{vanderVeer2026}
(\textbf{B}) Evolution of the vertical PFM phase (top panel) and the corresponding thin-film topography (bottom panel) of the WO\textsubscript{3} film under continuous scanning with an AC bias applied to the scanning probe tip. 
(\textbf{C}) Conducting AFM image acquired directly after the AC-bias-induced expansion of the domain walls (left) and after a \SI{12}{\hour} waiting period (right). 
The white dashed ellipses mark identical regions in the two cAFM images at the edge of the scanned area.
(\textbf{D}) Sketch of the domain-wall expansion, highlighting the three distinct regions observed in (C), along with a magnified view of the red-outlined area in (C).}
\label{fig:1_cAFM}
\end{figure}

% As our model system, we focus on thin films of $\text{WO}_3$.
Owing to its $A$-site vacant perovskite structure, known as the $\text{ReO}_{3}$ structure, $\text{WO}_{3}$ has the propensity to undergo a variety of oxygen octahedral tilts, as well as polar and anti-polar distortions that give rise to multiple first-order phase transitions between nearly-degenerate phases.~\cite{Salje1975,Hamdi2016,Mascello2020,Eckstein2022,Hassani2022}
The resulting flat energy landscape makes $\text{WO}_{3}$ an ideal system to investigate the interconvertibility between different phases.~\cite{Aird1998,Yun2020,Eckstein2024}
Here, we focus on epitaxial $\text{WO}_3$ films grown using pulsed laser deposition on (001)-oriented $\text{YAlO}_3$ (YAO) substrates (Materials and Methods).
In these films, the substrate-imposed epitaxial shear strain stabilizes a triclinic $P1$ phase (Fig.\ \ref{fig:1_cAFM}A) that is polar and features periodic in-plane polarized \SI{180}{\degree} stripe domains, as shown in the lateral piezoresponse force microscopy (PFM) image in the inset of Fig.\ \ref{fig:1_cAFM}A and reported previously.~\cite{vanderVeer2026}
In addition, the stripe-domain walls exhibit a reduction of all inherent structural tilts and distortions, giving rise to a higher symmetry and a locally enhanced conductivity via \emph{anti-distortive polaronic} charge accumulation.~\cite{vanderVeer2026,Hassani2025}.
Hence, our $\text{WO}_3$ films, prone to undergo first-order phase transitions, represent an ideal system with different symmetries of domain interior and domain wall that result in respective properties --- polarization and conductivity --- that are typically incompatible with one another in the same phase.\\

In these $\text{WO}_3$ films, upon continuous scanning with an AC bias ($\geq$\SI{2}{\volt}) applied to a scanning probe tip, we observe a gradual phase transition that emanates from the domain walls and clearly manifests both in the PFM signal and the thin-film topography (Figs.\ \ref{fig:1_cAFM}B, S1, S2).
In the vertical PFM phase (Fig.\ \ref{fig:1_cAFM}B, top panel), diamond-shaped regions with a near-zero phase first nucleate at the stripe-domain walls and progressively grow into extended domains as the scanning continues (Figs.\ S1, S2).
Simultaneously, trench-like topographical features that visibly increase the roughness of the scanned area (Fig.\ \ref{fig:1_cAFM}B, bottom panel) develop along well-defined crystallographic directions ($[110]_\text{YAO}$ and $[1\overline10]_\text{YAO}$) and delimit the diamond-shaped regions.
Conducting atomic force microscopy (cAFM) measurements performed directly after triggering this phase transition (Fig.\ \ref{fig:1_cAFM}C, left) reveal a markedly enhanced conductivity of the diamond-shaped domains.
In particular, as schematized in Fig.\ \ref{fig:1_cAFM}D, we can distinguish two main changes to the pristine domain configuration: 
(i) First, the confined domain walls expand into extended conducting domains. 
(ii) Within these expanded domain walls, a transition into an even more conducting phase occurs, which is separated from the domain-wall phase by a sharp boundary.
Strikingly, over the course of several hours in the absence of an external stimulus, the widened conducting domains progressively shrink until the pristine polar domain configuration with the confined conducting walls and the atomically-flat surface topography is restored (see Fig.\ \ref{fig:1_cAFM}C, right and Figs.\ S3, S4).
Thus, in these $\text{WO}_3$ films, applying an AC bias enables us to expand conducting domain walls into bulk domains in a metastable fashion, which gradually revert back to their pristine confined state within several hours. 
%%%%%%%%%%%%%%%%%%%%%%%%%%%%%%%%%%%%%%%%%%%%%%%%%%%%%%%%%%%%%%%%%%%%%%%%%%%%%%%%%%%%%%%%%%%
%End of Figure 1, Begin of Figure 2
%%%%%%%%%%%%%%%%%%%%%%%%%%%%%%%%%%%%%%%%%%%%%%%%%%%%%%%%%%%%%%%%%%%%%%%%%%%%%%%%%%%%%%%%%%%
\section*{Atomic-scale structure of the transient domain-wall phase}

\begin{figure}[ht!]
\includegraphics[width=0.98\columnwidth]{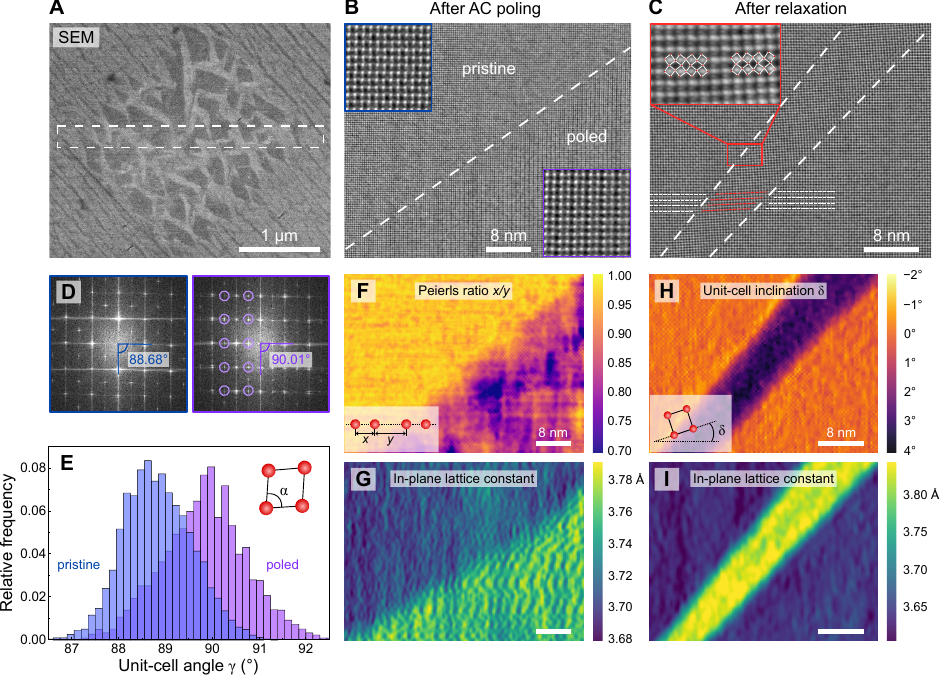}
\caption{\textbf{Atomic-scale structure of the expanded $\text{WO}_{3}$ domain walls.}
(\textbf{A}) Secondary electron contrast SEM image of a region with expanded domain walls appearing as dark diamonds and the pristine conducting domain walls as dark lines. 
The dashed outline marks the orientation of the extracted TEM lamella.
(\textbf{B}) Cross-sectional iDPC-STEM images in the [100] zone axis of $\text{WO}_{3}$ shortly after poling ($<$\SI{3}{\hour}).
The insets show magnified views of the top left (blue) and bottom right (purple) regions.
The dashed line indicates the phase boundary between pristine polar and expanded conducting domain-wall phase.
(\textbf{C}) iDPC-STEM image in the [100] zone axis of $\text{WO}_{3}$ after a waiting time of $\sim$\SI{12}{\hour}.
The diagonal dashed lines delimit a newly formed $\text{WO}_{3}$ twin domain.
The inset shows a magnified view across the twin domain wall.
The white and red dashed lines indicate the unit-cell inclination across the twin boundary.
(\textbf{D}) 2D FFT patterns extracted from (B) for the top left (blue) and bottom right (purple) regions.
(\textbf{E}) Histograms of the unit-cell angles $\alpha$ obtained from the HAADF-STEM image in Fig.\ S7 for the poled (purple) and the pristine region (blue).
(\textbf{F,G}) Maps of the horizontal Peierls distortion (F) and the horizontal lattice parameter (G) of the image in (B).
(\textbf{H,I})  Maps of the unit-cell inclination $\delta$ (H) and the horizontal lattice parameter (I) of the image in (C).
}
\label{fig:2_TEM}
\end{figure}

The cAFM image in Fig.\ \ref{fig:1_cAFM}C shows that the confined $\text{WO}_3$ domain walls expand, effectively introducing two new domain walls and a new intermediate domain, which in turn appears to become a nucleation site for a phase that is more conducting than that of the domain walls.
This raises the question on the nature of this highly-conducting, yet metastable phase of $\text{WO}_3$.
Therefore, to understand the atomic-scale structural changes giving rise to this transient phase within the expanded domain wall, we next move to cross-sectional scanning transmission electron microscopy (STEM) measurements on a poled region.
To achieve this, the induction of the phase transition by PFM, subsequent TEM lamella preparation using focused ion beam (FIB) milling, and STEM imaging are all done within \SI{3}{\hour}, such that domain-wall relaxation is minimal (Fig.\ S4).
% Doing so requires inducing the phase transition in PFM, preparing a TEM lamella using focused ion beam (FIB) milling, and STEM imaging, all within $3$-\SI{4}{\hour} and well before the domain-wall relaxation is complete (Fig.\ \ref{fig:SI_PFMrelaxation}).
During scanning electron microscopy (SEM) measurements, shown in Fig.\ \ref{fig:2_TEM}A, we can distinguish the phase-transition region, where the diamond-shaped conducting domains exhibit a strongly reduced secondary electron yield.
A similar suppression of the secondary electron emission is also evident at the confined domain walls of the as-grown film (see corners of the image).
This is expected for regions with increased conductivity in SEM, which tend to preferentially conduct away the primary electrons and, thus, result in a lower secondary electron emission.~\cite{vanderVeer2026,Aristov1984,Shih1997,Mosberg2019,vanRijn2024}\\

On this area, we extract a TEM lamella and perform integrated differential phase contrast (iDPC)-STEM imaging~\cite{Lazic2016} along the [100]\textsubscript{$\text{WO}_3$} (corresponding to $[1\overline{1}0]$\textsubscript{YAO}) zone axis of the $P1$ phase immediately after AC poling (Figs.\ \ref{fig:2_TEM}B, S5, S6) and after a prolonged waiting time (Fig.\ \ref{fig:2_TEM}C).
Directly after poling, shown in Fig.\ \ref{fig:2_TEM}B, a clear structural change in the oxygen sublattice is visible across a \SI{45}{\degree} inclined boundary (dashed line in Fig.\ \ref{fig:2_TEM}B).
The top left half features the characteristic anti-phase oxygen octahedral tilting along the vertical direction of the pristine triclinic phase~\cite{vanderVeer2026}, whereas on the bottom right half a periodicity doubling along the horizontal direction occurs.
In the two-dimensional fast Fourier transform (FFT) patterns derived from the pristine (Fig.\ \ref{fig:2_TEM}D, left) and poled (Fig.\ \ref{fig:2_TEM}D, right) regions, this periodicity doubling manifests as half-integer peaks (purple circles) in the horizontal direction in the right panel of Fig.\ \ref{fig:2_TEM}D.
Furthermore, mapping the horizontal bond disproportionation (or Peierls distortion) $x/y$ in real space~\cite{vanderVeerxx} (a value of $x/y\neq1$ indicates a deviation from equally spaced atoms), shown in Fig.\ \ref{fig:2_TEM}F, confirms the periodicity-doubled structure of the poled $\text{WO}_3$ phase.
The periodicity doubling originates entirely from the oxygen sublattice and does not appear in the corresponding high-angle annular dark field (HAADF)-STEM image (Fig.\ S7) that is sensitive only to the $\text{W}$ atomic positions.
The HAADF-STEM image instead reveals a horizontal lattice expansion of $1.2\pm$\SI{0.4}{\percent} in the poled phase, as shown in Figs.\ \ref{fig:2_TEM}G and S7.
This explains the vertical expansion and increase in roughness observed in the topography images in Fig.\ \ref{fig:1_cAFM}B.\\

Upon closer inspection of the unit-cell angle $\alpha$ in the poled and pristine regions in reciprocal (Figs.\ \ref{fig:2_TEM}D) and real space (Figs.\ \ref{fig:2_TEM}E, S7), we can further discern a change of $\alpha$ from the triclinic angle of $88.7\pm$\SI{0.7}{\degree} in the pristine phase, in line with previous XRD-based findings \cite{vanderVeer2026}, to $89.8\pm$\SI{0.8}{\degree} in the poled phase.
The same structural features, i.e.\ periodicity doubling and increase of the unit-cell angle to \SI{90}{\degree}, are also observed along the perpendicular [010]\textsubscript{$\text{WO}_3$} direction (Fig.\ S8).
Therefore, we can conclude that upon poling, the $\text{WO}_3$ undergoes a phase transition from its pristine triclinic to a (pseudo)tetragonal structure that nucleates at the location of the domain walls.
This transition may proceed through intermediate monoclinic and/or orthorhombic phases, akin to bulk $\text{WO}_3$, for which the STEM characterization, however, does not provide clear evidence.\\

After a sufficient waiting time to allow the expanded domain walls to relax (see Fig.\ \ref{fig:1_cAFM}C), the STEM images in Figs.\ \ref{fig:2_TEM}C and S9 reveal that the periodicity-doubled tetragonal phase has completely disappeared.
Instead, a triclinic needle domain with an inclined unit cell (see Fig.\ \ref{fig:2_TEM}H) and a \SIadj{5}{\percent}-expanded horizontal lattice constant (see Fig.\ \ref{fig:2_TEM}I) with respect to the pristine phase has formed at the former triclinic/tetragonal phase boundary.\\

To assess the $\text{WO}_{3}$ domain-wall expansion and its underlying structural phase transition on a macroscopic level, we next perform Raman spectroscopy measurements on both pristine and poled $\text{WO}_{3}$ regions, as shown in Figs.\ \ref{fig:3_Raman_COMSOL}A, \ref{fig:3_Raman_COMSOL}B, S10, and S11.
As a highly symmetry-sensitive probe, Raman scattering constitutes an ideal technique to further investigate the subtle structural changes observed in STEM in Fig.\ \ref{fig:2_TEM}.~\cite{Seo2023}
The Raman spectra in the high-frequency range in Fig.\ \ref{fig:3_Raman_COMSOL}B, predominantly governed by modes involving the light oxygen atoms, indeed show pronounced spectral differences between pristine and poled areas that can be rigorously separated from substrate signals by non-negative matrix factorization (NMF) \cite{Kunc2013,Woelffel2015}, shown in Fig.\ S12.
The pristine film exhibits two main intense features at \SI{708}{\per\centi\meter} and \SI{806}{\per\centi\meter} that are also present in the $P2_{1}/n$ and $P2_{1}/c$ phases and can be assigned to the anti-polar motion of oxygen atoms along $y$ and $z$, respectively.~\cite{Hassani2022}
For the poled region, in contrast, the band at \SI{708}{\per\centi\meter} becomes markedly suppressed, while the previous shoulder at \SI{672}{\per\centi\meter} develops into a distinct feature.
The change from two to only one high-frequency Raman band in the $700-$\SI{850}{\per\centi\meter} range is qualitatively consistent with DFT-calculated Raman spectra of the bulk tetragonal $\text{WO}_{3}$ phases $P4/ncc$ and $P4/nmm$ (see Fig.\ S13), in which only the $M_{3}^{-}$ mode corresponding to an anti-polar distortion along $z$ remains.
Crucially, these higher-symmetry phases in $\text{WO}_{3}$ feature a significantly reduced bandgap ($<$\SI{2}{\eV}) compared to low-symmetry ones ($>$\SI{3}{\eV})~\cite{Hassani2025}, which likely accounts for the appearance of the poled areas as dark squares in the optical microscopy image in the inset of Fig.\ \ref{fig:3_Raman_COMSOL}A.
It is further consistent with the increased conductivity in these areas observed in cAFM in Fig.\ \ref{fig:1_cAFM}C.\\
%%%%%%%%%%%%%%%%%%%%%%%%%%%%%%%%%%%%%%%%%%%%%%%%%%%%%%%%%%%%%%%%%%%%%%%%%%%%%%%%%%%%%%%%%%%
%End of Figure 2, Begin of Figure 3
%%%%%%%%%%%%%%%%%%%%%%%%%%%%%%%%%%%%%%%%%%%%%%%%%%%%%%%%%%%%%%%%%%%%%%%%%%%%%%%%%%%%%%%%%%%
\section*{Driving force for domain-wall-to-domain conversion}

\begin{figure}[htbp]
\includegraphics[width=\columnwidth]{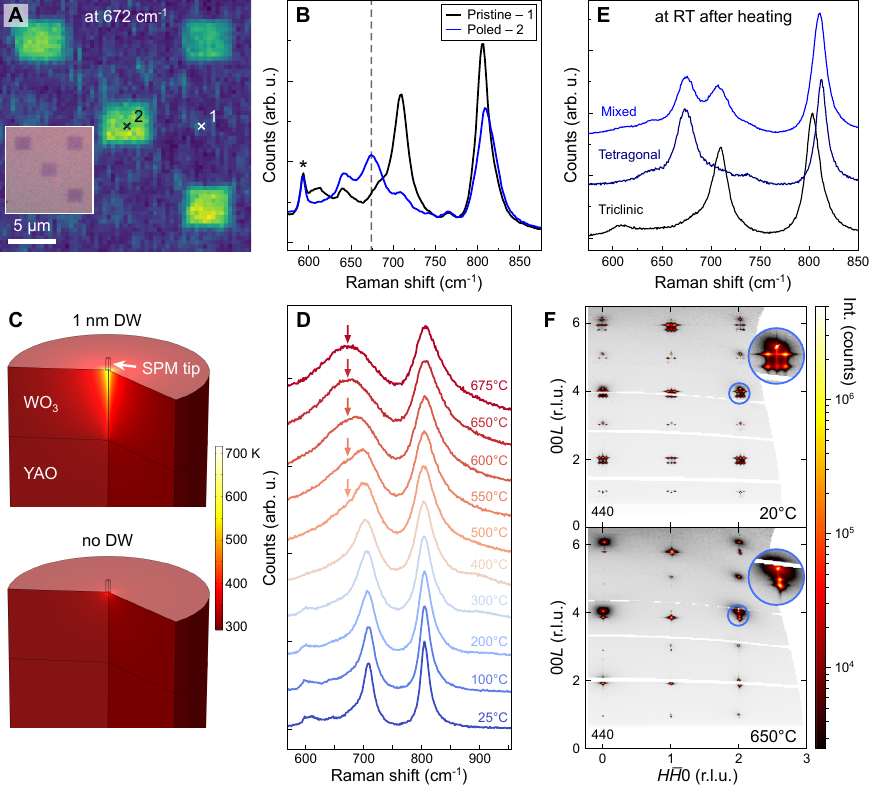}
\caption{\textbf{Thermally-induced domain-wall expansion in $\text{WO}_{3}$.} 
(\textbf{A,B}) Comparative Raman spectroscopy analysis of as-grown $\text{WO}_{3}$  and expanded conducting domain walls. 
(A) Two-dimensional Raman map at a wavenumber of \SI{672}{\per\centi\meter} across an area containing four $5\times$\SI{5}{\micro\meter\squared} poled regions.
The inset shows the white light optical microscope image of the same area.
(B) Raman spectra of the pristine and poled regions marked in (A).
The asterisk marks a YAO substrate peak and the dashed line indicates the location of the map in (A).
(\textbf{C}) Finite-element simulations of the temperature profile across the $\text{WO}_{3}$ film in the absence (bottom) and presence (top) of a \SI{1}{\nano\meter} wide domain wall, assuming a \SI{5}{\nano\meter} tip contact radius, a \SI{4}{\volt} AC amplitude, and a \SI{400}{\kilo\hertz} frequency.
(\textbf{D}) Raman spectra acquired at different temperatures during heating.
(\textbf{E}) Raman spectra collected at different sample positions, as marked in Fig.\ S19, at room temperature after prior sample heating to \SI{675}{\celsius}.
(\textbf{F}) $(HH0)$ reciprocal-space reconstructions with origin 440 in the frame of the triclinic $\text{WO}_{3}$ phase, measured at room temperature (top) and at \SI{650}{\celsius} (bottom).
The insets shows enlarged views of the $624$ reflections.
}
\label{fig:3_Raman_COMSOL}
\end{figure}

While both the atomic-scale STEM (Fig.\ \ref{fig:2_TEM}) and the macroscopic Raman characterization point toward a phase transition from triclinic to (pseudo)tetragonal at the root of the conducting domain wall expansion, in bulk $\text{WO}_{3}$, the tetragonal phases only become stable at elevated temperatures above \SI{775}{\celsius}.~\cite{Eckstein2022}
Similarly, triclinic epitaxial $\text{WO}_{3}$ films transition into their non-polar parent phase above \SI{550}{\celsius}.~\cite{vanderVeer2026}
This suggests that strongly localized heating of the conducting domain walls via the AC electric field of the scanning probe tip is the trigger for the phase transition and the resulting domain-wall expansion.
Notably, tip pressure alone (Fig.\ S14) and DC electric fields (Fig.\ S15) do not induce the domain-wall expansion, whereas AC electric fields, independent of the frequency (see Fig.\ S16), do.\\

To test whether an AC electric field can induce such a temperature increase, we perform finite-element modeling of the voltage and temperature profiles across the $\text{WO}_{3}$ film induced by an AC electric field applied to a PFM tip (Figs.\ \ref{fig:3_Raman_COMSOL}C, S17, and S18).
The calculations assume a radius of the tip-sample contact area of \SI{5}{\nano\meter}~\cite{Lu2012} and an AC electric field amplitude of \SI{4}{\volt} at \SI{400}{\kilo\hertz}, as also used experimentally (see Methods section for details of the modeling).
The temperature profiles, shown in Fig.\ \ref{fig:3_Raman_COMSOL}C, in the presence (top) and absence (bottom) of a \SIadj{1}{\nano\meter}-wide conducting domain wall indeed reveal strongly localized heating up to at least \SI{700}{\kelvin} around the conducting domain wall, while the bulk of the film undergoes insignificant heating.
Moreover, the calculations show that the heating is substantially lower for a DC or quasi-DC bias (Fig.\ S18A) and largely independent of the AC bias frequency, (Fig.\ S18B,C), consistent with our findings from PFM (Figs.\ S15, S16).
This difference in heating between AC and DC voltages can be attributed to the presence of the substrate acting as a series capacitor.
Under DC biases, the current flow is limited to capacitive charging.
With an AC bias applied, the capacitive reactance drops, which lowers the series impedance, allows for a larger current, and subsequently increases the energy dissipation via heating.
Temperature-dependent Raman spectroscopy measurements, shown in Fig.\ \ref{fig:3_Raman_COMSOL}D, further support local heating as the main driving force for the phase transition and the domain-wall expansion.
Upon heating above \SI{550}{\celsius}, a shoulder feature appears in the spectra at \SI{672}{\per\centi\meter} and progressively grows at the expense of the band at \SI{708}{\per\centi\meter} that eventually disappears entirely within the thermal broadening at higher temperatures.
This evolution is in good agreement with the spectral differences observed between the poled and pristine Raman spectra in Fig.\ \ref{fig:3_Raman_COMSOL}B.
Interestingly, after cooling to room temperature, we find that the phase transition is not entirely reversible, as evidenced by the Raman spectra acquired at different sample locations (Figs.\ \ref{fig:3_Raman_COMSOL}E, S19).
The majority of the film exhibits the typical Raman spectrum of the pristine triclinic phase.
Other regions that are also visible in an optical microscope (Fig.\ S19), however, keep the Raman signatures of the high-temperature tetragonal phase or that of a triclinic/tetragonal phase mixture.
In a very similar manner, we also observe that poled regions at room temperature retain permanent spectral changes (Fig.\ S20) despite an apparent complete relaxation at the film surface (Fig.\ S4).
These remanent changes may originate from local plastic deformations, as suggested by the STEM measurements (Fig.\ \ref{fig:2_TEM}C) after relaxation, resulting in different compatibility constraints of the starting and final states.\\

To remove any remaining ambiguity on the structure of the high-temperature $\text{WO}_{3}$ phase in our films, we further employ synchrotron-based large-area 3D reciprocal space surveys.~\cite{Bang2025,Zatterin2024,vanGentGonzalez2026}
The $(HH0)$ reciprocal lattice reconstructions, shown in Fig.\ \ref{fig:3_Raman_COMSOL}F, at room temperature (top) and at \SI{650}{\celsius} (bottom) along with perpendicular $(H\overline{H}0)$ reconstructions (Fig.\ S21) verify a phase transition from triclinic to tetragonal and allow us to exclude potential monoclinic, orthorhombic, or cubic phases.
Specifically, the four-fold peak splitting characteristic of the triclinic phase~\cite{vanderVeer2026}, visible in the magnified view in Fig.\ \ref{fig:3_Raman_COMSOL}F around the $624$ reflection, completely disappears at \SI{650}{\celsius}, consistent with a transformation from triclinic to tetragonal.\\

Hence, the direct agreement between the structural changes in the $\text{WO}_{3}$ from STEM and Raman under tip-induced AC poling, as well as Raman and XRD measurements under macroscopic sample heating, allow us to conclude that tip-induced Joule heating constitutes the primary driving force for (i) the domain-wall expansion and (ii) the subsequent transition from triclinic to tetragonal structures occurring within the expanded domain walls.
%%%%%%%%%%%%%%%%%%%%%%%%%%%%%%%%%%%%%%%%%%%%%%%%%%%%%%%%%%%%%%%%%%%%%%%%%%%%%%%%%%%%%%%%%%%
%End of Figure 3, Begin of Figure 4
%%%%%%%%%%%%%%%%%%%%%%%%%%%%%%%%%%%%%%%%%%%%%%%%%%%%%%%%%%%%%%%%%%%%%%%%%%%%%%%%%%%%%%%%%%%
\section*{Volatile synaptic switching in $\text{WO}_{3}$ films}

The possibility to electrically trigger this phase transition from polar triclinic to the highly conducting tetragonal parent phase at room temperature and thereby expand the conducting $\text{WO}_{3}$ domain walls provides a direct handle on the local electrical properties.
This may be attractive for device applications.
One particularly intriguing aspect in this regard is our ability to tune the phase fractions of the triclinic and tetragonal phases (see Fig.\ \ref{fig:3_Raman_COMSOL}E, S20) with low and high electrical conductivities, respectively, in a continuous manner due to the first-order phase transition.\\

\begin{figure}[ht!]
\includegraphics[width=\columnwidth]{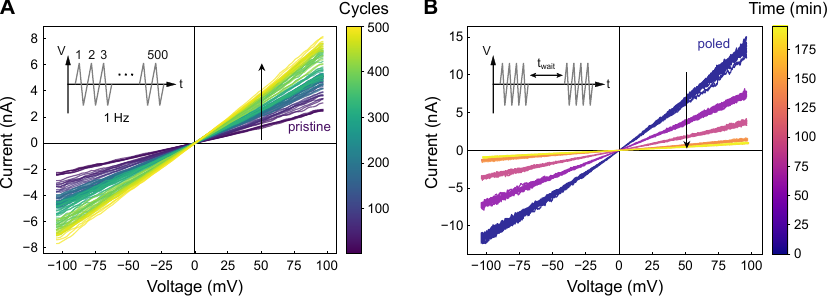}
\caption{\textbf{Synaptic modulation of film conductivity via continuous domain-wall-to-domain conversion.} 
(\textbf{A}) $I$-$V$ characteristics measured at room temperature with a cAFM tip on a pristine region of the $\text{WO}_{3}$ film.
The $IV$ sweeps are obtained under the application of a triangular waveform at \SI{1}{\hertz} with $V_\text{max}$ values of $\pm$\SI{100}{\milli\volt} and 500 cycles.
(\textbf{B}) $I$-$V$ characteristics measured at room temperature with a cAFM tip on a PFM-poled region of the $\text{WO}_{3}$ film.
For each time step, 10 cycles of a triangular voltage pulse with $V_\text{max}$ of $\pm$\SI{100}{\milli\volt} at \SI{1}{\hertz} was applied.
}
\label{fig:4_memristive}
\end{figure}

To investigate the potential of the domain-wall-to-domain interconversion as the basis for synaptic devices with tunable electrical resistances, we measure the $I$-$V$ characteristics of the $\text{WO}_{3}$ film locally using a conductive cAFM tip (Fig.\ \ref{fig:4_memristive}).
When placing the cAFM tip statically on a pristine region of the $\text{WO}_{3}$ film and applying cyclical triangular voltage pulses, the obtained  $I$-$V$ measurements (Fig.\ \ref{fig:4_memristive}A) reveal a quasi-continuous increase in the measured current with increasing cycle number.
If, in turn, we place the cAFM tip onto a previously expanded conducting domain and track the evolution of the $I$-$V$ characteristics over time (Fig.\ \ref{fig:4_memristive}B), we observe a slow but continuous conductivity decrease with a saturation and a complete recovery of the conductivity of the pristine film after a couple of hours.
The time scale of the conductivity recovery in Fig.\ \ref{fig:4_memristive}B closely matches that of the phase relaxation from tetragonal to triclinic (Fig.\ S20) and of the recovery of the confined conducting domain walls (Fig.\ S4).\\

Such an analog tunability of the $\text{WO}_{3}$ resistivity, exploiting the interconversion between a polar pristine and a conducting domain-wall phase, has the potential for applications as volatile synaptic units, in which the local electrical properties can be precisely and continuously tuned at the nanoscale.
It may also be suitable for physical reservoir computing, which requires complex and tunable input-output characteristics, as well as relaxation of its memory state.~\cite{Everschor-Sitte2024}

%%%%%%%%%%%%%%%%%%%%%%%%%%%%%%%%%%%%%%%%%%%%%%%%%%%%%%%%%%%%%%%%%%%%%%%%%%%%%%%%%%%%%%%%%%%
%End of Figure 4
%%%%%%%%%%%%%%%%%%%%%%%%%%%%%%%%%%%%%%%%%%%%%%%%%%%%%%%%%%%%%%%%%%%%%%%%%%%%%%%%%%%%%%%%%%%
\section*{Conclusion}
Owing to their ability to be written, moved, and erased, domain walls with unique functional properties in ferroic materials that are symmetry-forbidden in bulk have inspired multiple device applications over the past 15 years and given rise to the field of domain-wall nanoelectronics.~\cite{Catalan2012}
Ferroelectric materials with an insulating bulk and conducting domain walls have been particularly attractive for their ability to accommodate the otherwise incompatible properties of long-range polar order and metallic conductivity.~\cite{Anderson1965}
So far, however, domain-wall properties have been exclusively regarded as being confined to a fixed two-dimensional geometry without the possibility to modify their spatial dimension.\\
Our results demonstrate the first realization of such an interconversion between two-dimensional conducting domain walls and three-dimensional conducting domains within polar epitaxial $\text{WO}_{3}$ films at room temperature.
The key principle enabling this interconversion is the strongly localized selective heating of the conducting $\text{WO}_{3}$ domain walls by means of an AC electric field, which induces a first-order phase transition to the non-polar tetragonal phase.
Accessing this phase transition proceeds through a widening of the domain walls that subsequently act as nucleation sites for the highly conducting tetragonal phase.
By exploiting the first-order nature of this phase transition and the resulting phase coexistence, we further accomplish a continuous tunability of the local electrical conductivity, highlighting the potential of domain-wall-to-domain conversion for neuromorphic applications and nanoelectronics at large.\\
We believe that this presented approach to interconvert conducting domain walls and polar domains may likely be extensible to other material systems, provided some key conditions are met:
(i) A higher-energy phase with a larger conductivity (lower bandgap) has to exist at experimentally accessible temperatures. Notably, this coincides with the condition for materials showing anti-distortive polarons.~\cite{Hassani2025}
(ii) The high-temperature phase needs to be accessible via a first-order phase transition.
(iii) Strain compatibility between the low and high-temperature phases must be met, to enable the pre-existence of the high-temperature phase within a domain wall away from the phase transition temperature.

%%%%%%%%%%%%%%%%%%%%%%%%%%%%%%%%%%%%%%%%%%%%%%%%%%%%%%%%%%%%%%%%%%%%%%%%%%%%%%%%%%%%%%%%%%%
%End of Main Text
%%%%%%%%%%%%%%%%%%%%%%%%%%%%%%%%%%%%%%%%%%%%%%%%%%%%%%%%%%%%%%%%%%%%%%%%%%%%%%%%%%%%%%%%%%%
\newpage
\begin{acknowledgments}
\noindent 
The authors thank Jacob Baas and Joost Zoestbergen for technical support.
We acknowledge the European Synchrotron Radiation Facility (ESRF) for provision of synchrotron radiation facilities under proposal number HC-6505.
We acknowledge support from the Zernike Institute for Advanced Materials (ZIAM) and the Groningen Cognitive Systems and Materials Center (CogniGron) for the electron microscopy facility at the University of Groningen.
We thank Charles McMonagle for the assistance during the synchrotron measurements, Wasim Akram and Guillaume Nataf for preliminary thermal conductivity measurements, and Mael Guennou and Mohammad Shoker for the support with Raman measurements and Raman facility access.
\end{acknowledgments}

\section*{Funding}
\noindent 
M.F.S.\ acknowledges funding from the HORIZON-MSCA-2024-PF RECOMPUTE (101203197)
M.F.S., M.A., J.vG.G., H.R.J.C., M.A.L., B.J.K., and B.N.\ acknowledge financial support from the Groningen Cognitive Systems and Materials Center (CogniGron) and the Ubbo Emmius Foundation of the University of Groningen.
M.F.S., H.R.J.C., and B.N.\ acknowledge funding from the HORIZON-CL4-2023 grant CONCEPT (101135946).
J.P. and M.A.L. acknowledge funding from the European Union via the ERC Advanced Grant DEOM (101055097).
E.vdV.\ acknowledges financial support by the Research Center Future Energy Materials and Systems (RC FEMS) of the Research Alliance Ruhr.
M.C.W.\ and T.P.\ acknowledge financial support by the R\'egion des Pays de la Loire through the Etoiles Montantes project No. 2022\_11808.
A.S.\ and P.G.\ acknowledge financial support from the Fonds de la Recherche Scientifique (FNRS) through the PDR project ToPoTex (T.0128.25) and access to the CECI supercomputing facilities funded by the FNRS (2.5020.1), as well as to the Tier-1 supercomputer of the Walloon Region Lucia (1910247).

\section*{Author contributions}
\noindent 
M.F.S.\ conceived the project and coordinated it jointly with B.N. 
M.F.S.\ performed the epitaxial thin-film growth, laboratory-scale XRD, AFM, PFM, and cAFM measurements.
M.A.\ performed the STEM sample preparation and STEM acquisition.
STEM data analysis was performed by E.vdV.\ together with M.A., A.S.G., and M.F.S.
Raman data acquisition was performed by M.C.W.\ and J.P.\ together with M.F.S.
Raman data analysis was done by M.C.W.\ with support by T.P.
Finite-element simulations were performed by J.vG.G.
DFT calculations were performed by H.H.\ and A.S.
Synchrotron-based SCXRD measurements were performed by M.F.S., J.vG.G., E.vdV., and D.A.C.
Reciprocal-space reconstructions were done by J.vG.G.
cAFM and macroscopic electrical measurements were performed by M.F.S.\ with support from H.R.J.C.
Project supervision was done by M.A.L., P.G., B.J.K., and B.N.
All authors participated in the analysis, discussion, and interpretation of the results.
M.F.S.\ and B.N.\ wrote the manuscript with inputs from all authors.
\section*{Competing financial interests}
\noindent The authors declare no conflicts of interest.
\section*{Data availability}
\noindent Data collected at the European Synchrotron Radiation Facility as part of proposal HC-6505 is available at the following DOI: 10.15151/ESRF-ES-2403714702.\\
Further data supporting this study are available in DataverseNL at tbd.

% \clearpage
% \begin{figure}[ht!]
% \includegraphics[width=\columnwidth]{Figures_maintext/Fig1_V2.pdf}
% \caption{\textbf{Caption title.} Caption}
% \label{fig:1}
% \end{figure}

\clearpage


\section*{Supplementary material}

# Materials and Methods

**Thin-film growth**. The $WO_3$ films were grown on single-crystalline (001)-oriented $YAlO_3$ substrates (space group *Pbnm*, Crystec GmbH) by pulsed laser deposition using a 248 nm KrF excimer laser (Coherent Corp.). The films were grown from a sintered ceramic $WO_3$ target at a substrate temperature of 680 °C, a laser fluence of $0.5\,\mathrm{J\,cm^{-2}}$ (spot size: $2.1\,\mathrm{mm^2}$), an oxygen partial pressure of 0.1 mbar, a laser frequency of 2 Hz, and a target-substrate distance of 50 mm. Before cooldown at $10\,^{\circ}\mathrm{C\,s^{-1}}$, the films were annealed at 680 °C in growth pressure for 1 h. In total, 8000 pulses were deposited, resulting in a final $WO_3$ film thickness of 150 nm, as confirmed by in-situ reflection high-energy electron diffraction and ex-situ X-ray reflectivity measurements.

**Scanning probe microscopy**. Atomic force microscopy measurements were performed on a Bruker Dimension Icon scanning probe microscopy in tapping mode using Budget Sensors Tap300Al-G tips ($k = 40\,\mathrm{N\,m^{-1}}$). Piezoresponse force microscopy measurements were carried out on an Asylum Research Cypher ES atomic force microscope with PtIr-coated tips (Bruker SCM-PIC-V2, $k = 0.1\,\mathrm{N\,m^{-1}}$ and Budget Sensors Multi75E-G, $k = 3\,\mathrm{N\,m^{-1}}$). To acquire lateral and vertical PFM images, the long axis of the cantilever was aligned along the [100] direction of the YAO substrate. All PFM measurements were performed sub-resonance with an approximate tip-sample force of 15 nN. cAFM measurements were carried out on the same instrument using the softer SCM-PIC-V2 probes. Conducting silver paint was placed on the surface of the film to which electrical biases of 100-1200 mV were applied. Subsequently, the current flowing into the grounded tip was measured during scanning.

**X-ray diffraction**. Symmetric $\theta$-$2\theta$ measurements were performed in line focus on a Panalytical X'Pert MRD thin-film diffractometer, equipped with a Cu source (Cu-K$\alpha$, $\lambda = 1.54\,\text{Å}$), a 2xGe(220) hybrid monochromator, and a PIXcel$^{3D}$ area detector.
Large-scale 3D reciprocal space surveys were carried out at the BM01 beamline at the European Synchrotron Radiation Facility (ESRF) equipped with Pilatus 2M detectors. The measurements were performed at an incident wavelength of 1 Å (12.4 keV) to minimize fluorescence from film and substrate, and improve the film sensitivity. Data acquisition was performed in shutterless mode and an incident $\omega$ angle of 5° with continuous rotation of the sample and consisted of 180° scan with integration over an angular range of 0.1° for 1.5 s. Temperature-dependent measurements were carried out in the same manner on a custom heating stage in air. Raw data frames were processed with the CrysAlisPro suite to retrieve the UB matrix, which was subsequently used to produce 2D reconstructions of reciprocal space using Python.

**Scanning electron microscopy (SEM) measurements and focused ion beam (FIB) sample preparation**. SEM measurements and electron-transparent STEM sample fabrication were performed using an FEI Helios G5 CX dual-beam SEM-FIB instrument. To facilitate relocating a previously PFM-poled region in SEM, the following measures were taken: (1) $2 \times 2\,\mu\mathrm{m}^2$ and $5 \times 5\,\mu\mathrm{m}^2$ areas were poled in very close proximity to distinctive topographical features. (2) Silver paint on the

film surface and direct sample contact with a grounding needle (EasyLift) a few µm away from the imaged area were utilized to reduce charging. (3) A low acceleration voltage of 1 kV was used to reduce the penetration depth of the primary electron beam. (4) The sample was electrically isolated from the stage by mounting it on a sapphire disk.

After relocating the region of interest, protective EBID carbon (C) and platinum (Pt) layers were deposited before the IBID Pt protective layer. Cross-sectional chunks with dimensions of $14 \times 2 \times 5$ µm$^3$ were cut and transferred using an EasyLift needle to a copper half-grid. The sample was then initially thinned down to 80-100 nm in thickness using Ga-beam processing at 30 kV on an opening window with dimensions of $6 \times 5$ µm$^2$. The remaining chunk was left thick, maintaining a rigid frame to minimize bending and stress release in the e-transparent window. Finally, several low kV cleaning steps (5 and 2 kV) were used to clean the side surfaces of the lamella.

Since the PFM-poled regions restored back to the pristine state within <6 h (see Fig. S4), PFM poling, SEM imaging, FIB lamella preparation, and TEM imaging were all performed within <4 h.

**Scanning transmission electron microscopy**. The FIB-prepared lamella on the TEM grid was transferred to the TEM column using a low-background double-tilt TEM holder optimized to collect X-rays. The STEM measurements were performed with a double abberration-corrected (probe and image correctors) and monochromated Themis Z scanning transmission electron microscope (Thermo Fisher Scientific) operating at 300 kV. The STEM images were acquired in the HAADF (high-angle annular dark-field) and iDPC (integrated differential phase contrast) imaging modes. A beam convergence angle of ∼25 mrad and a probe current of 30−50 pA were used for STEM imaging.

**Raman spectroscopy**.

Two Raman spectrometers were used for this study. A) An InVia Qontor confocal Raman microscope (Renishaw, UK) using a Leica 50x N Plan objective (0.50NA) to focus the output of a 532 nm continuous wave laser onto the sample. A Renishaw Edge filter was used to block the Rayleigh scattering (<100 cm$^{-1}$), which was further suppressed with a pinhole. The Raman scattered light was then dispersed by a 1800 lines/mm grating and imaged by a CCD camera. B) An InVia Renishaw Reflex Raman Microscope (Renishaw, UK) in micro-Raman mode with a spectral cut off at 70 cm$^{-1}$, spectral resolution was provided by a 2400 lines/mm grating. For excitation a 633 nm He-Ne laser and a 532 nm diode laser were used. Spatial resolution was provided by a Leica 100x and a 50x long working distance objective for room temperature and temperature-dependent measurements, respectively. Great care was taken to prevent sample heating. No laser power induced band shift were observed for the used laser powers For all the measurements, the [100] direction of the $YAlO_3$ substrate was aligned parallel to the laser polarization plane. Single Raman spectra were collected in high-confocality mode with a laser power on the sample of approximately 40 mW by integrating for 10 s in each acquisition. To account for the background signal of the $YAlO_3$ substrate, spectra were acquired on both pristine substrates and by focusing the optical excitation deep (>10 µm) into the substrate yielding comparable results. For the long-term relaxation tracking of

the poled phase, single spectra were acquired on the same spot every 5 min overnight or until the relaxation was complete. Raman maps were collected over $25 \times 25\,\mu m^2$ areas with a step size of 500 nm and same acquisition parameters as for the single spectra. For temperature dependent measurements a Linkam TS1500 stage was used.

**Raman data analysis: Separation of thin-film and substrate contribution**. Thin-film Raman spectra typically contain overlapping contributions from the film and the substrate. To separate these contributions, we combined depth-scan Raman spectroscopy with non-negative matrix factorization (NMF), a multivariate method that represents a set of measured spectra as combinations of a small number of non-negative component spectra [1, 2, 3, 4]. During a depth scan, the focal plane is moved through the sample, so that the Raman signal is collected from different depths. When the focal plane is within the substrate, the substrate contribution becomes dominant. In contrast, the thin-film contribution is mainly localized near the sample surface. Consequently, the film and substrate contributions vary differently as a function of depth, which allows them to be separated by NMF. Let $Y$ denote the matrix containing all Raman spectra measured during a depth scan. NMF approximates this matrix as the product of two non-negative matrices,

$$Y \approx CS,\ C \text{ and } S \geq 0. \tag{1}$$

The columns of the component matrix $C$ represent the extracted spectral components, while the rows of the score matrix $S$ describe the contribution of these components at the different positions of the depth scan. In the present case, two components were used: one associated primarily with the thin film and one associated primarily with the substrate. The NMF decomposition is obtained by minimizing an objective function that combines two requirements: the reconstructed spectra $CS$ should reproduce the measured spectra $Y$, and the elements of the matrices $C$ and $S$ should not become unnecessarily large. For the present analysis, the objective function was:

$$\begin{aligned} L(C,S) &= 0.5 * ||Y - CS||^2_{loss} \\ &\quad + 0.5 * \alpha_C * n_{features} * ||C||^2_{Frobenius} \\ &\quad + 0.5 * \alpha_S * n_{samples} * ||S||^2_{Frobenius} \end{aligned}$$

The first term is the reconstruction error and measures the difference between the measured spectra and those reconstructed from the NMF components. The second term penalizes large values in the component spectra, while the third term penalizes large component amplitudes across the depth scan. The parameters $\alpha_C$ and $\alpha_S$ control the strength of these two regularization terms. The quantities $n_{\text{features}}$ and $n_{\text{samples}}$ denote the number of spectral data points per spectrum and the number of spectra in the depth scan, respectively. Thus, the algorithm searches for non-negative matrices $C$ and $S$ that reproduce the experimental data while avoiding unnecessarily large matrix elements. Two components were selected based on the expected physical contributions from the substrate and the thin film. The regularization parameters and initialization procedure were selected for each dataset to obtain a physically meaningful separation. Typical depth scans extended from 3 µm above the thin film to 10 µm into the substrate, with a step size of 200–300 nm.

A linear baseline was fitted to and subtracted from each spectrum before applying NMF. The thin-film spectrum shown in the main text corresponds to the component identified as the thin-film contribution by the NMF analysis. All data processing was performed based on the Python NMF implementation provided by scikit-learn [5].

**DFT calculations**.
First-principles density functional theory (DFT) calculations were performed using the Abinit package. [6, 7] The interactions between ionic core and valence electrons are modeled using a norm-conserving (NC) pseudo-potential. [8] Exchange-correlation energy functional is approximated using local density of electrons (LDA) and parametrized within the Perdew-Wang 1992 (PW92) scheme. [9, 10] The Kohn-Sham (KS) wavefunctions are represented using a plane-wave basis truncated at a kinetic energy cutoff of 70 Ha. The Brillouin zones of $P4/ncc$ and $P4/nmm$ phases are sampled in a uniform grid of 6x6x4 and 6x6x8 mesh of k-points. Numerical solution of the KS equation is obtained iteratively until the wavefunction residuals of the occupied states are converged within $10^{-22}$. The responses such as phonon frequency at the $\Gamma$ point and Raman activity of modes are obtained using density functional perturbation theory (DFPT). Note, here Raman spectra of the tetragonal phases are calculated for $WO_3$ films clamped to $LaAlO_3$, which slightly shifts the peak positions with respect to those obtained in the corresponding bulk phases. [11]

**Finite-element modeling**.
Finite element modeling of the electric field applied by an SPM tip and resulting heating of $WO_3$ is performed using a two-dimensional axisymmetric model and the electromagnetic heating multiphysics module in COMSOL Multiphysics. [12] The film thickness, domain-wall size, and SPM tip size all follow those in the corresponding experiments. Relevant material properties for the $WO_3$ film and domain walls are shown in Tab. S1 and kept the same across both types of $WO_3$ with the exception of the electrical conductivity, which is set based on our resistance measurements across (for bulk $WO_3$ film) and parallel to the $WO_3$ domain walls. [13]

| **Property** | **Bulk $WO_3$** | **$WO_3$ domain wall** | **Reference** |
|---|---|---|---|
| Thermal conductivity | $2.5\,W\,m^{-1}\,K^{-1}$ | $2.5\,W\,m^{-1}\,K^{-1}$ | preliminary measurements by W. Akram & G. Nataf |
| Electrical conductivity | $33\,S\,m^{-1}$ | $1\times10^{3}\,S\,m^{-1}$ | [13] |
| Relative permittivity | 40 | 40 | [14, 15] |
| Heat capacity | $32\,J\,kg^{-1}\,K^{-1}$ | $32\,J\,kg^{-1}\,K^{-1}$ | [16] |
| Density | $7.16\,g\,cm^{-3}$ | $7.16\,g\,cm^{-3}$ | XRR fitting |

Table S1: $WO_3$ materials properties used for the COMSOL simulations.

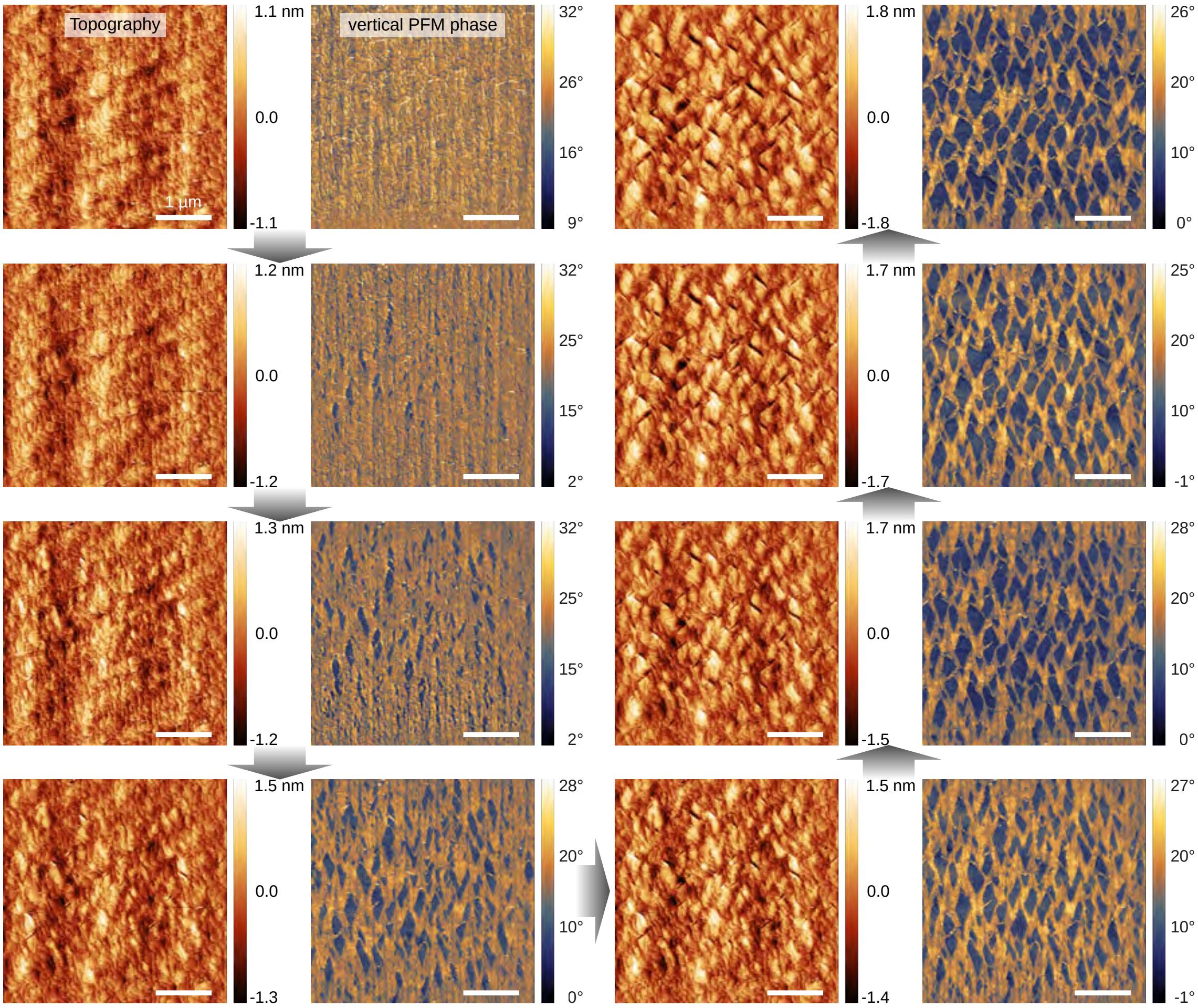


Figure S1: **Detailed evolution of the domain-wall expansion.** Evolution of the thin films topography (left) and the vertical PFM phase (right) under continuous scanning of a pristine $WO_3$ region with an AC bias of 2 V applied to the scanning probe tip. The duration of each consecutive scan is ∼10 min. In the pristine films (top left), the in-plane polarized stripe domains are faintly visible in both the topography and the vertical PFM phase. With increasing scanning duration, diamond-shaped areas nucleate at the stripe-domain walls, expand, and occasionally merge. The thin-film topography simultaneously develops trench-like features.

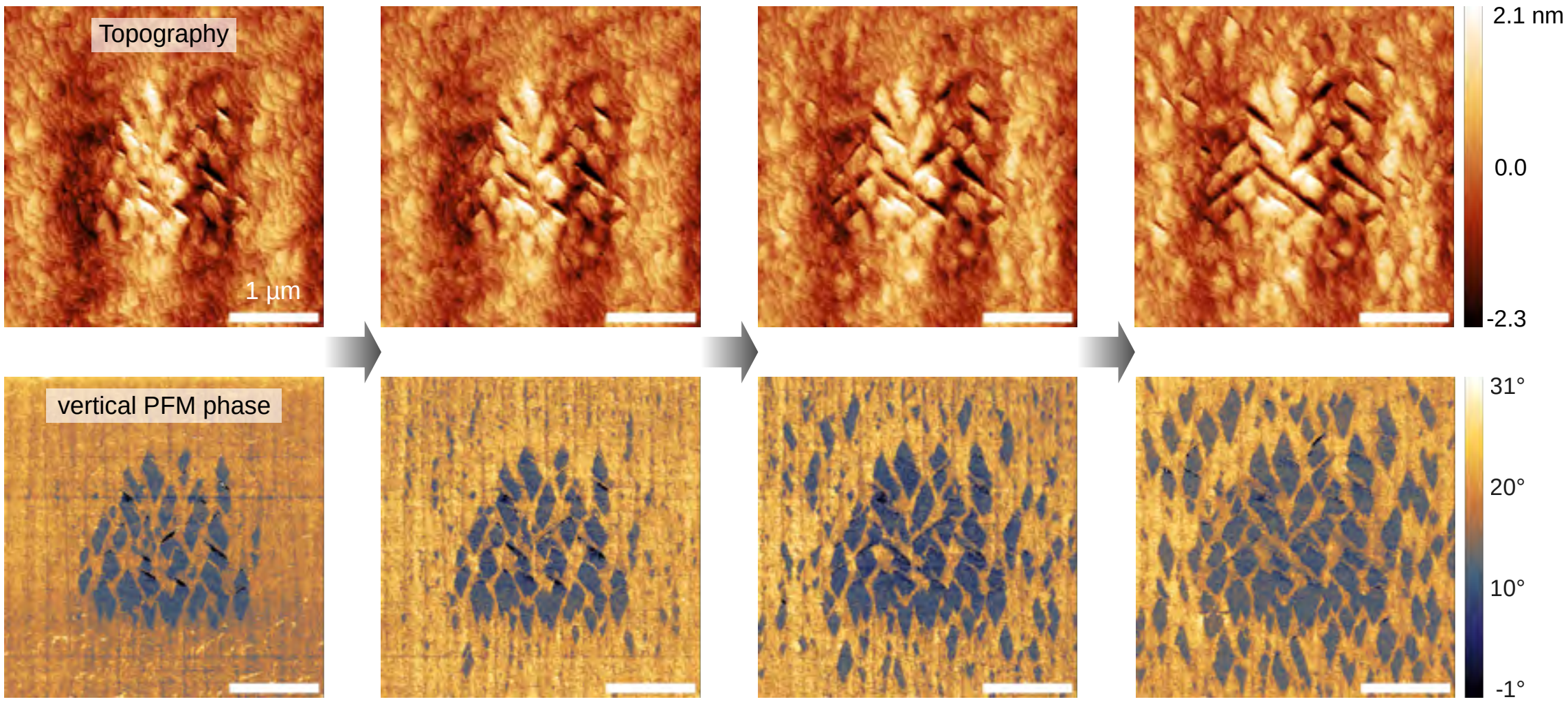


Figure S2: **Domain-wall expansion across a pre-poled area.** Upon continuous scanning with an AC bias applied to the scanning probe tip across a previously poled region, we observe the same domain-wall expansion as in Fig. S1. While new diamond-shaped expanded domain walls appear in the pristine films, the pre-existing domains continue to expand and merge.

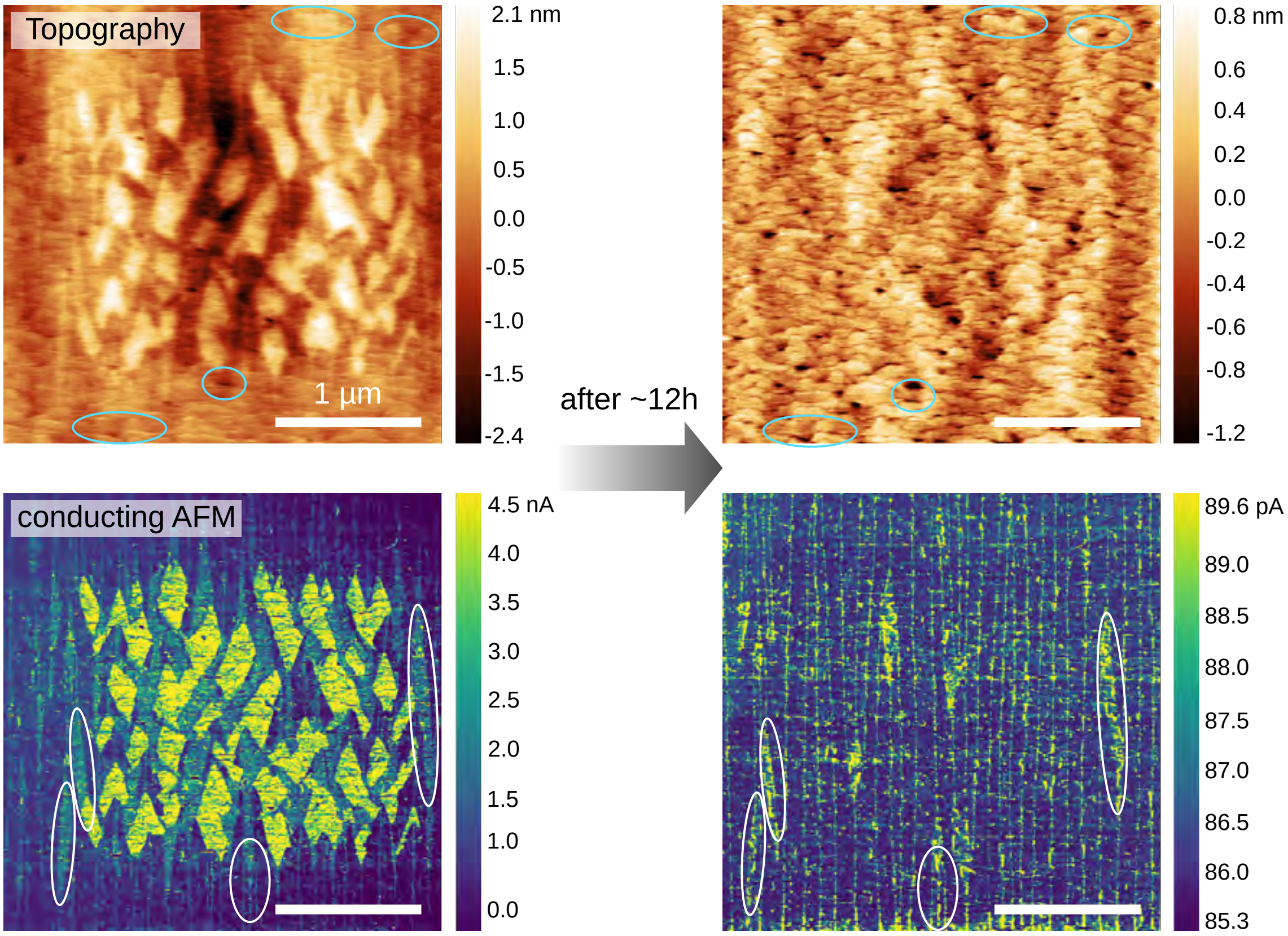


Figure S3: **Restoration of the confined domain walls and the pristine thin-film topography.** Conducting AFM image and thin-film topography acquired directly after domain-wall expansion (left) and after a waiting period of ∼12 h without continuous scanning (right). Directly after poling, the expanded conducting domain walls manifest clearly as diamond-shaped domains both in cAFM and the topography. After the waiting time, the pristine terraced topography is perfectly restored, along with the confined conducting domain walls. The outlined white and blue ellipses serve as a guide to the eye to indicate identical regions before/after domain-wall restoration in cAFM and the thin-film topography, respectively.

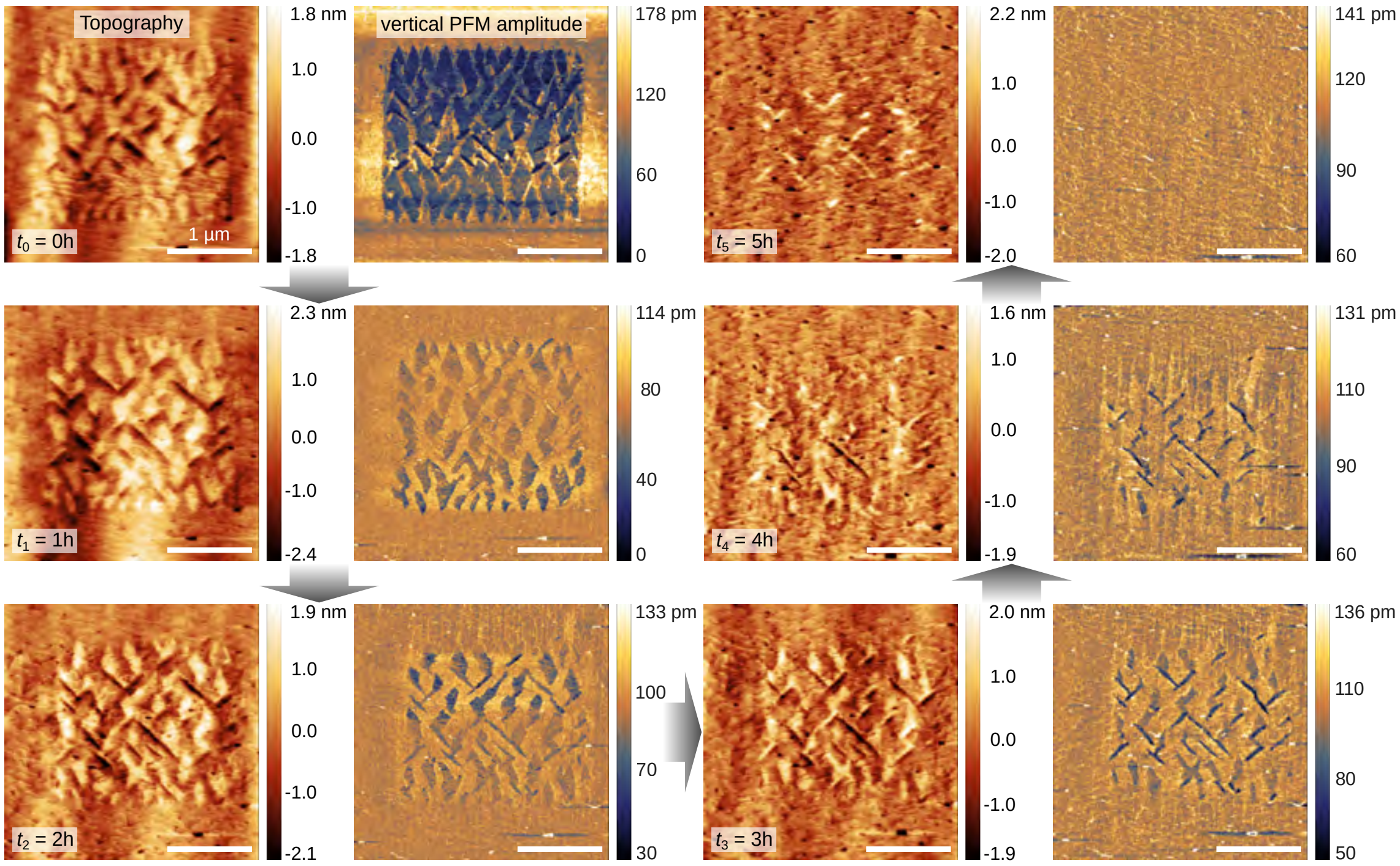


Figure S4: **Detailed evolution of the domain-wall restoration process.** Evolution of the thin-film topography (left) and the vertical PFM amplitude (right) while waiting for the relaxation of the expanded conducting domains into confined domain walls. The PFM scans are acquired with an AC bias of 1 V that does not by itself induce the domain-wall expansion. Between consecutive scans, there is waiting period of 1 h, while each scan takes ~10 min. With longer waiting times, the expanded diamond-shaped domains shrink continuously, until they completely disappear after ~5 h. Similarly, the domain-shaped elevated regions in the topography disappear, as well as the trenches delimiting these regions. After 5 h only a few scar-like topographical features remain at the original location of the trenches.

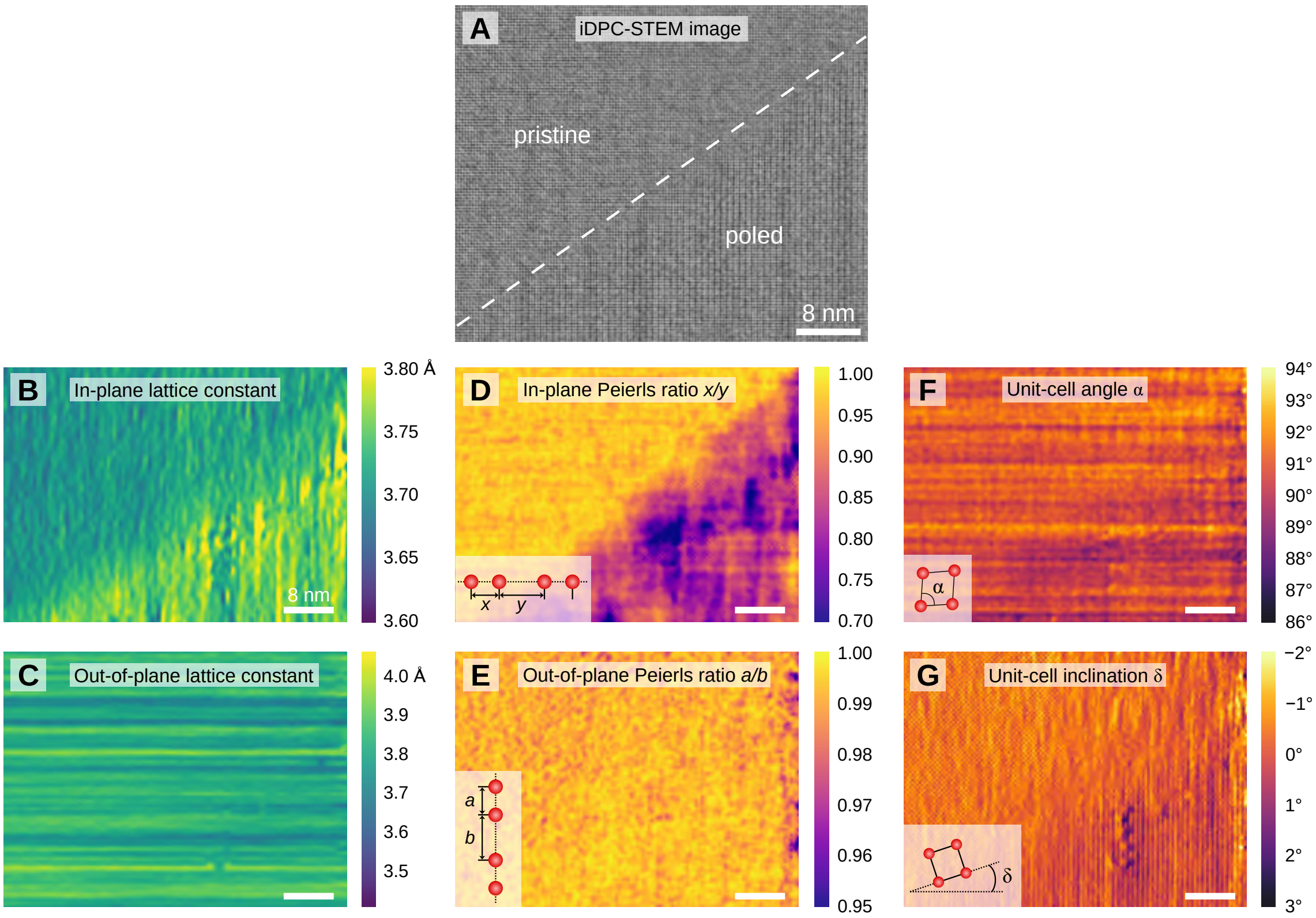


Figure S5: **Additional atomic-structure iDPC-STEM data of the expanded $WO_3$ domain wall.** (**A**) iDPC-STEM image in the [100] zone axis of $WO_3$, as also shown in Fig. 2B. (**B,C**) Maps of the horizontal (B) and vertical (C) lattice parameters obtained from (A). (**D,E**) Maps of the horizontal (D) and vertical (E) Peierls distortions obtained from (A), revealing a pronounced horizontal unit-cell doubling in (D) that is absent in Fig. S7D. (**F,G**) Maps of the unit-cell angle $\alpha$ (F) and the unit-cell inclination $\delta$ (G) obtained from (A).

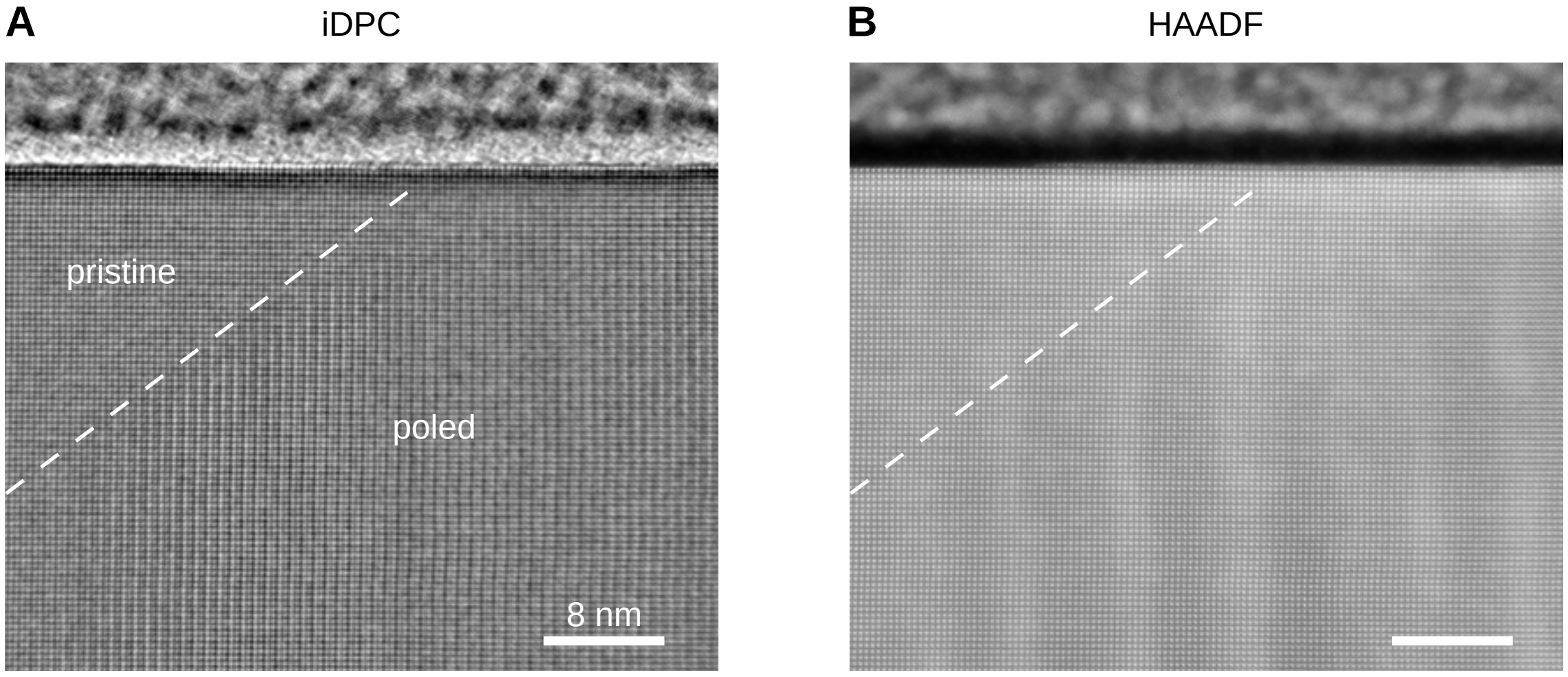


Figure S6: **Additional atomic-structure STEM images of the expanded $WO_3$ domain walls in the [100] zone of $WO_3$.** (**A,B**) iDPC-STEM image (A) and HAADF-STEM image (B) in the [100] zone axis of $WO_3$ for a poled region reaching the film surface. The periodicity doubled structure with a tetragonal unit-cell angle of 90° reaches the surface of the film.

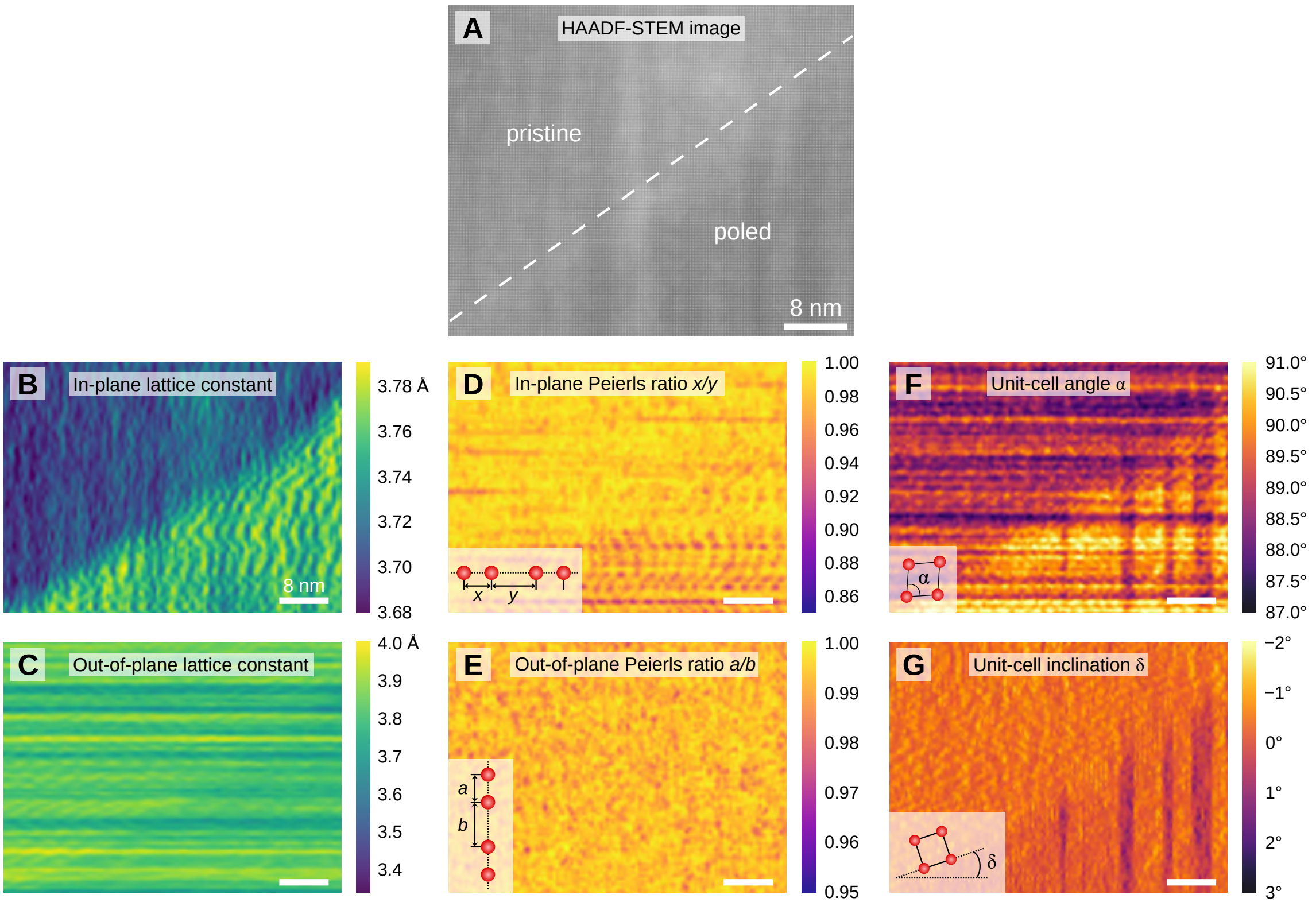


Figure S7: **Additional atomic-structure HAADF-STEM data of the expanded $WO_3$ domain wall.** (**A**) HAADF-STEM image in the [100] zone axis of $WO_3$ of the same region as shown in Fig. 2B. The diagonal dashed line marks the boundary between the poled (bottom right) and pristine (top left) regions. (**B,C**) Maps of the horizontal (B) and vertical (C) lattice parameters obtained from (A). Whereas the vertical lattice parameter does not show any abrupt variations across the phase boundary, the horizontal lattice constant is expanded by 1.1 % in the poled phase. (**D,E**) Maps of the horizontal (D) and vertical (E) Peierls distortions obtained from (A), revealing no unit-cell doubling in the W sublattice. Therefore, the unit-cell doubling, seen in iDPC-STEM image in Fig. 2B, originates entirely from the oxygen sublattice. (**F,G**) Maps of the unit-cell angle $\alpha$ (F) and the unit-cell inclination $\delta$ (G) obtained from (A). In the poled region, the unit-cell angle gets close to 90°, which points to the emergence of a tetragonal $WO_3$ structure. The histograms of the unit-cell angles in Fig. 2E were obtained by extracting the local unit-cell angles in (F).

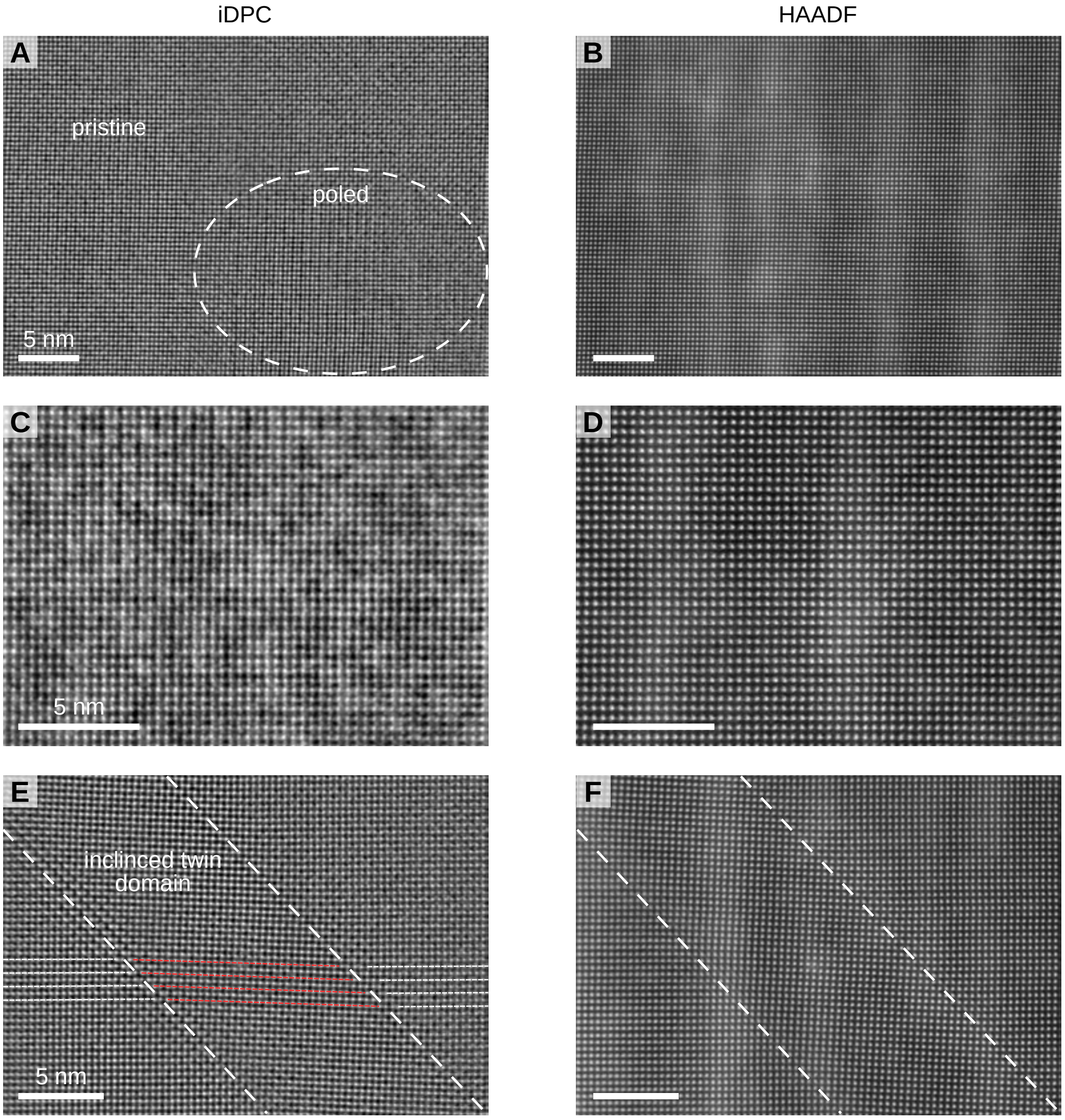


Figure S8: **Additional atomic-structure STEM images of the expanded $WO_3$ domain walls in the [010] zone of $WO_3$.** (**A,B**) iDPC-STEM image (A) and HAADF-STEM image (B) in the [010] zone axis of $WO_3$. A region with horizontal periodicity doubling, as observed similarly in Fig. 2B, can be seen within the dashed outline of the iDPC image. The atomic structure of the poled region in the [010] zone axis, shown here, is equivalent to the one in the [100] zone axis in Fig. 2B, which allows us to conclude on a tetragonal structure of the poled $WO_3$. Note that for this lamella in the [010] zone, the poled region is less clearly separated from the pristine structure than in Fig. 2B, likely due to a longer delay time between PFM poling, FIB lamella preparation, and STEM imaging (total delay time $\sim$ 5-6 h) (**C,D**) Magnified iDPC-STEM (C) and HAADF-STEM (D) images in the [010] zone axis of $WO_3$ of a poled region exhibiting the periodicity doubling. (**E,F**) iDPC-STEM (E) and HAADF-STEM (F) images in the [010] zone axis of $WO_3$ after complete relaxation of the poled areas.

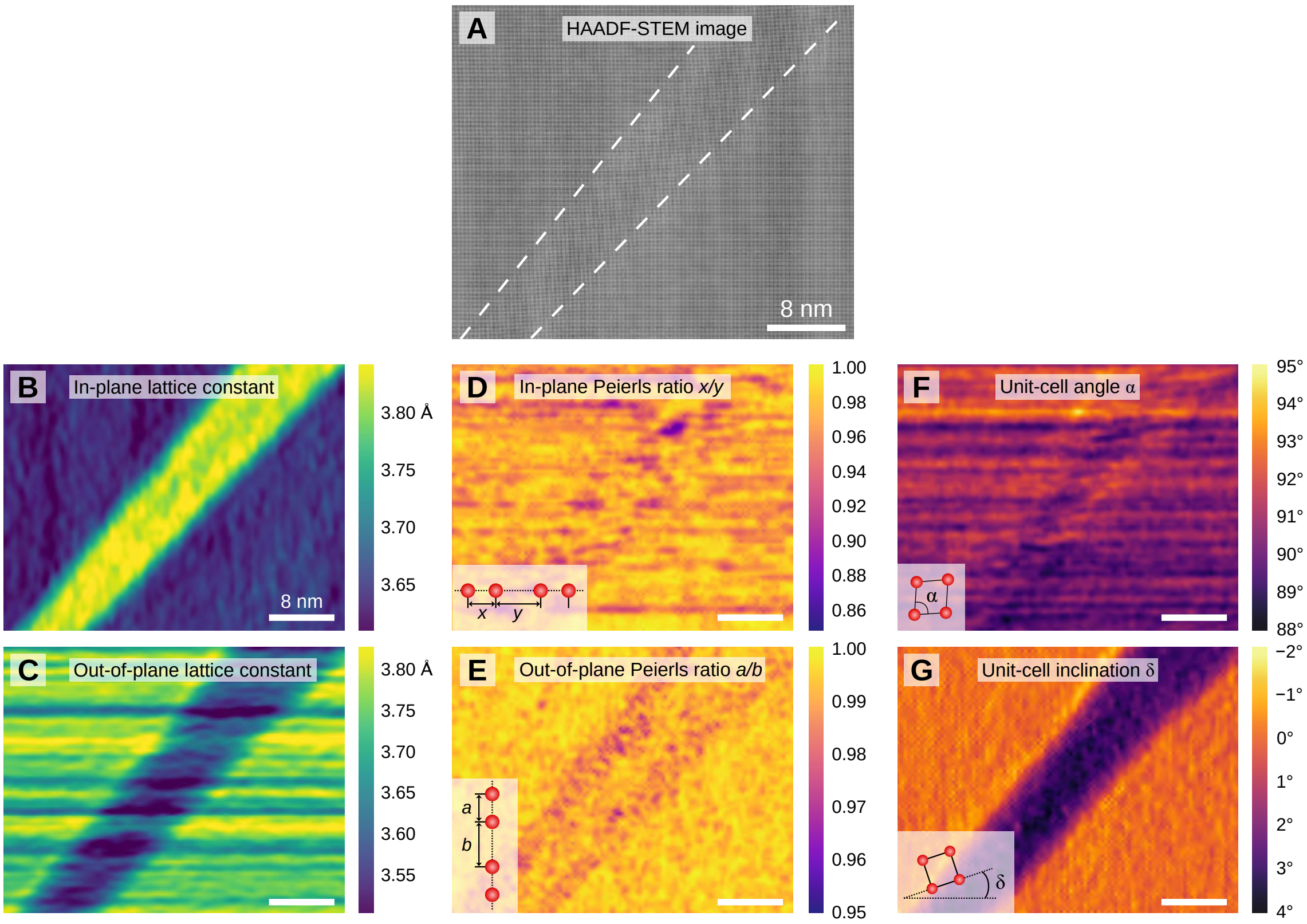


Figure S9: **Additional atomic-structure HAADF-STEM data after relaxation of the expanded $WO_3$ domain walls.** (**A**) HAADF-STEM image in the [100] zone axis of $WO_3$ of the same region as shown in Fig. 2H. (**B,C**) Maps of the horizontal (B) and vertical (C) lattice parameters obtained from (A). The inclined $WO_3$ twin-domain needle exhibits a horizontal lattice expansion of 5.1 % along with a reduction of the vertical lattice parameter by −4.3 %. (**D,E**) Maps of the horizontal (D) and vertical (E) Peierls distortions obtained from (A). No clear unit-cell doubling is evident. (**F,G**) Maps of the unit-cell angle $\alpha$ (F) and the unit-cell inclination $\delta$ (G) obtained from (A). Within the twin domain, the triclinic unit cell is inclined by about −2.6°.

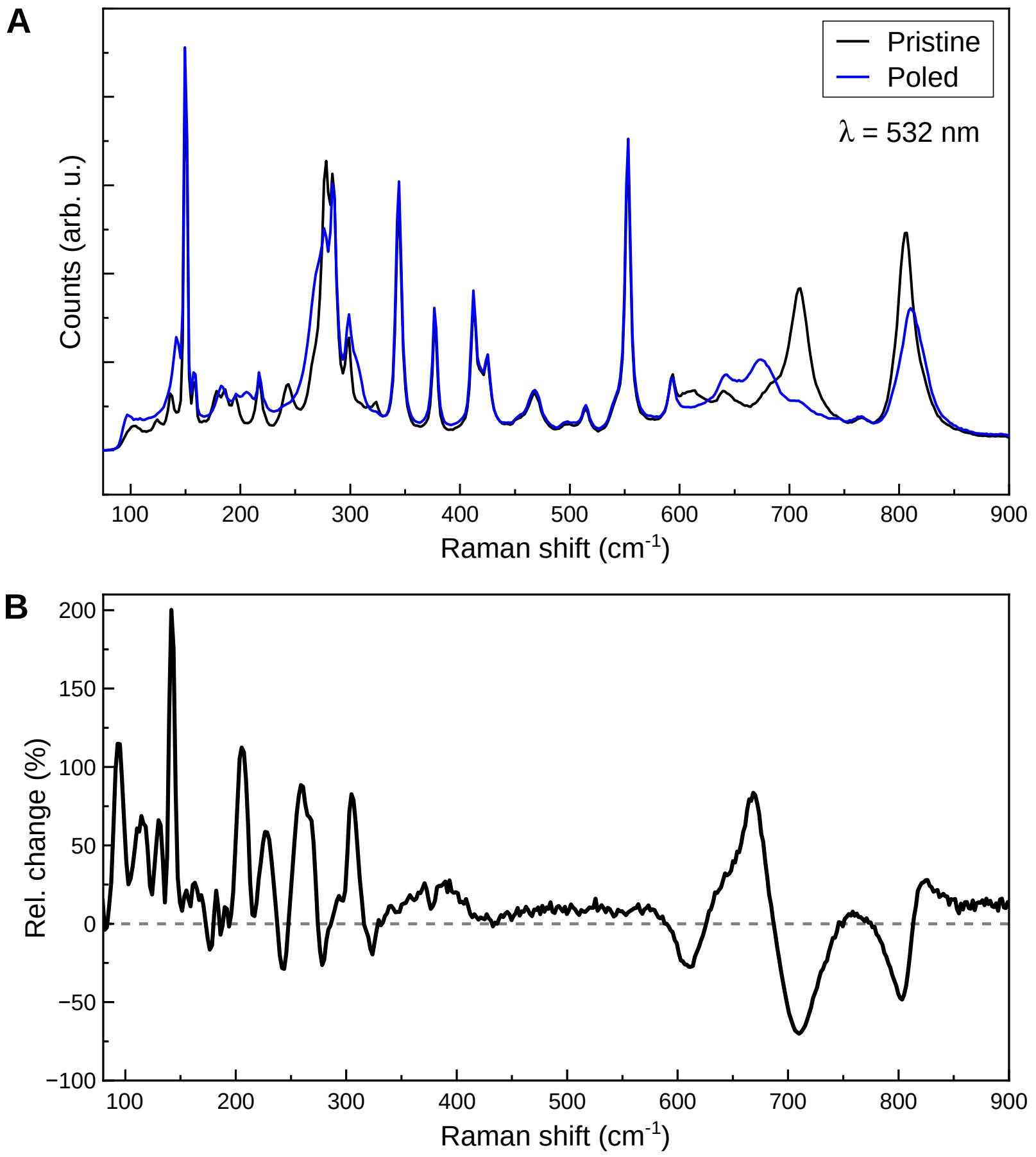


Figure S10: **Full-range Raman spectra of pristine and poled $WO_3$.** (**A**) Complete raw Raman spectra of pristine and poled $WO_3$ acquired with an excitation wavelength of 532 nm without NMF into substrate and film components. The spectral region between $320\,cm^{-1}$ and $600\,cm^{-1}$ is dominated by $YAlO_3$ substrate contributions. Fig. 3B shows a zoom in of the high-frequency range of these spectra. (**B**) Relative difference of the poled and pristine Raman signals across the full spectral range, highlighting the dominant substrate contributions in the mid-frequency range and the most pronounced changes between pristine (triclinic) and poled (tetragonal) in the low- and high-frequency regimes.

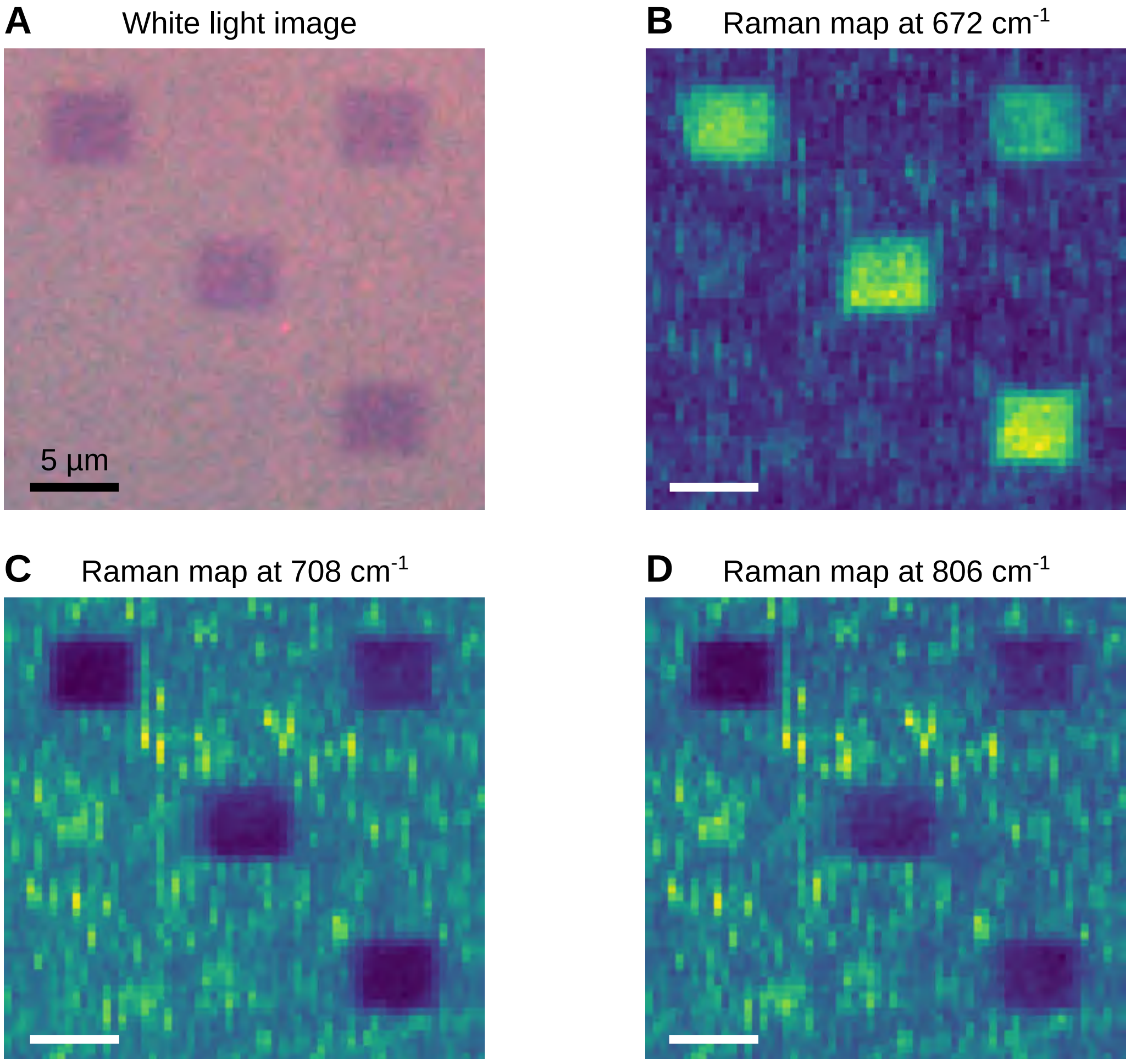


Figure S11: **Additional two-dimensional Raman maps across poled $WO_3$ regions.** (**A**) White light optical microscope image of the investigated area containing four 5×5 µm$^2$ poled regions appearing as dark squares. (**B-D**) Two-dimensional maps displaying the Raman intensity at the most intense features in the high-frequency range of the poled tetragonal phase at 672 cm$^{-1}$ (B), and of the pristine triclinic phase at 708 cm$^{-1}$ (C) and 806 cm$^{-1}$ (D), respectively.

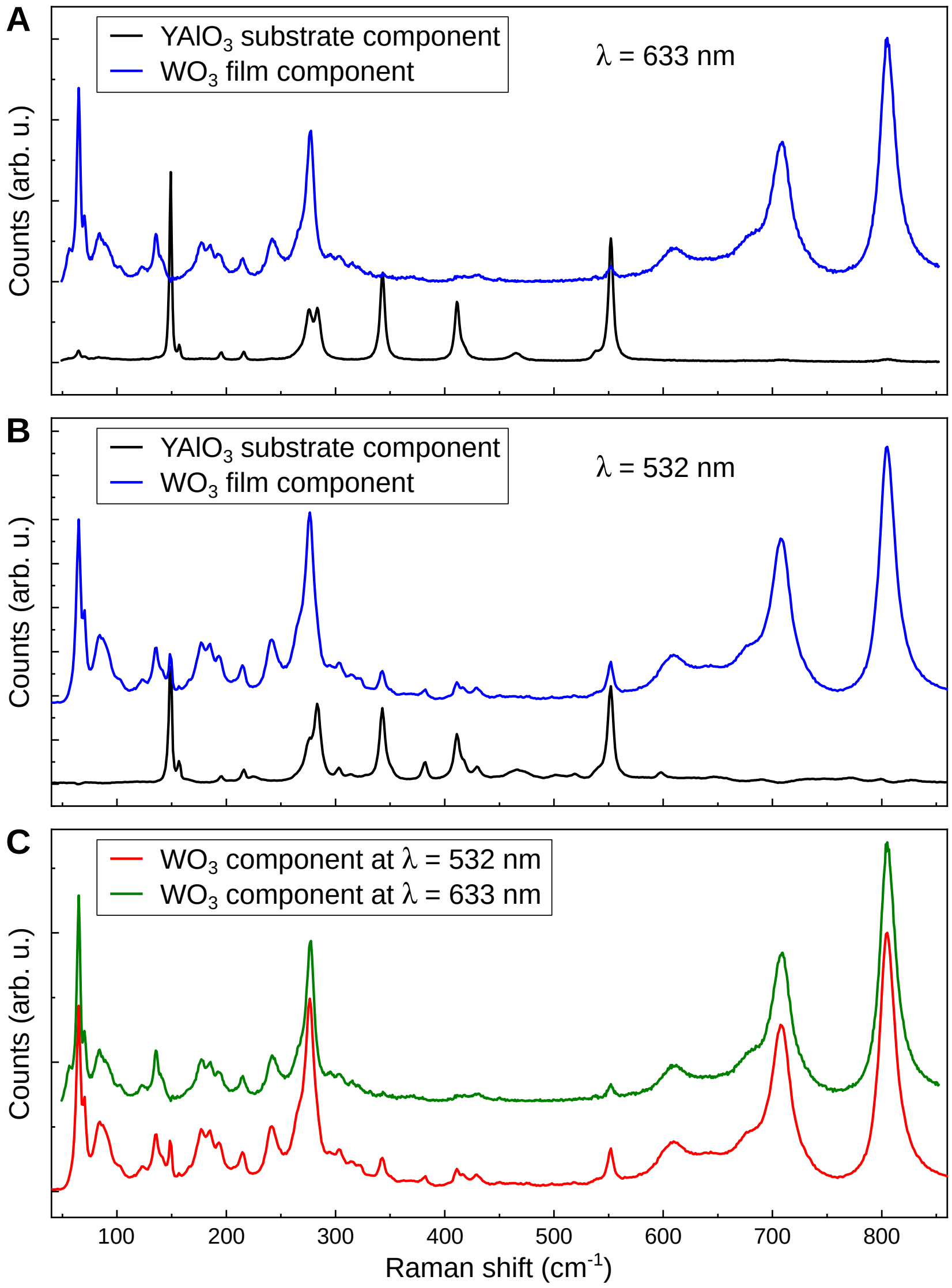


Figure S12: **Non-negative matrix factorization (NMF) of Raman spectra.** (**A,B**) NMF components originating from the $YAlO_3$ substrate and the $WO_3$ film for Raman spectra acquired with excitation wavelengths of (A) 633 nm and (B) 532 nm. The substrate components were identified by acquiring a series of Raman spectra at varying focal depths of the excitation laser spots. The strong suppression or complete absence of substrate spectral features in the $WO_3$ spectra indicates a successful spectral decomposition. Note that the difference of the substrate contributions for the two excitation wavelengths can be attributed to the presence of photoluminescence of the $YAlO_3$ substrate under 532 nm excitation. (**C**) Comparison of the Raman spectra for the two excitation wavelengths. Both spectra exhibit identical spectral features with only minor remaining signals from the substrate. In particular, the spectra in the high-frequency regime (>600 $cm^{-1}$) are perfectly identical without any spurious substrate contributions.

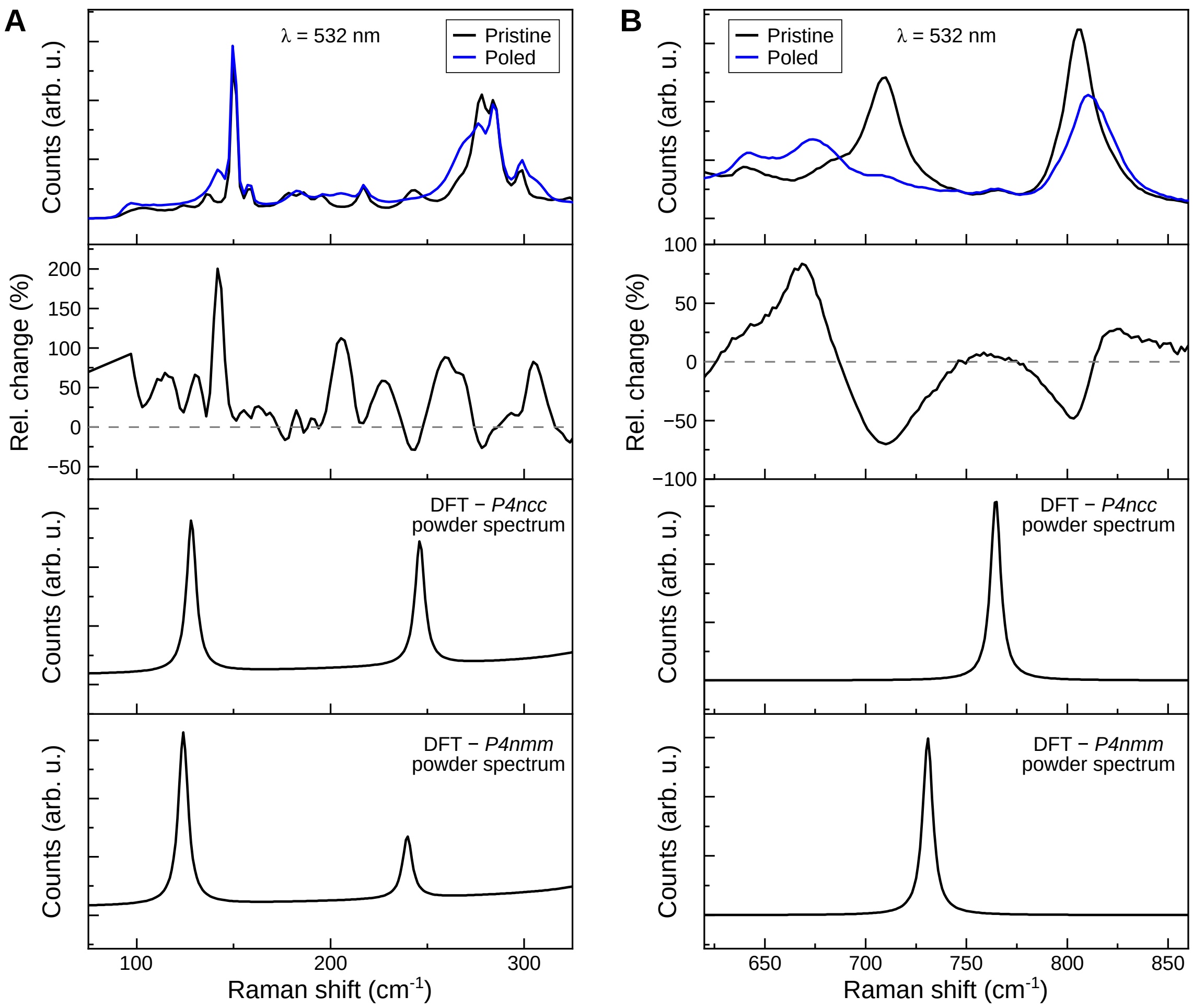


Figure S13: **Comparison of experimental and DFT-calculated Raman spectra.** (**A,B**) Comparison of the (A) low-frequency and (B) high-frequency spectral regions of experimental and DFT-calculated Raman spectra. The Raman spectra are acquired with a laser excitation wavelength of 532 nm. The middle panel shows the relative spectral differences between pristine and poled areas of the $WO_3$ film. The bottom two panels show the DFT-calculated powder Raman spectra at 0 K for the two tetragonal phases of $WO_3$, i.e. $P4/ncc$ and $P4/nmm$, that are stable in bulk crystals. [17] In the high-frequency range, the tetragonal phases are characterized by only one intense peak originating from a $M_3^-$ mode, corresponding to an anti-polar oxygen octahedral distortion along z.

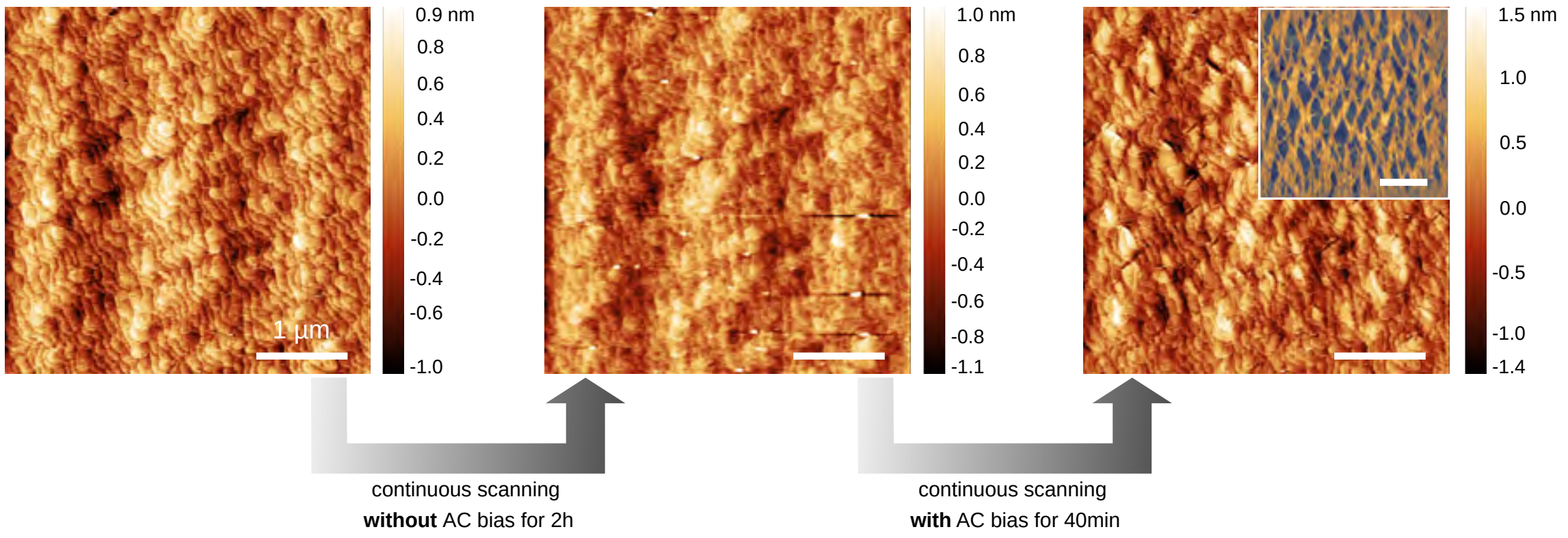


Figure S14: **Effect of tip pressure on the domain-wall expansion in $WO_3$.** Under continuous scanning with a tip pressure of approximately 15 nN and no AC bias for 2 h, no changes in thin-film topography associated with the domain-wall-to-domain expansion can be discerned. When subsequently reapplying an AC bias of $\geq$2 V, the domain-wall expansion occurs again within 40 min and manifests both in the thin-film topography as diagonal trench-like features and the vertical PFM phase as diamond-shaped domains (inset of top right image).

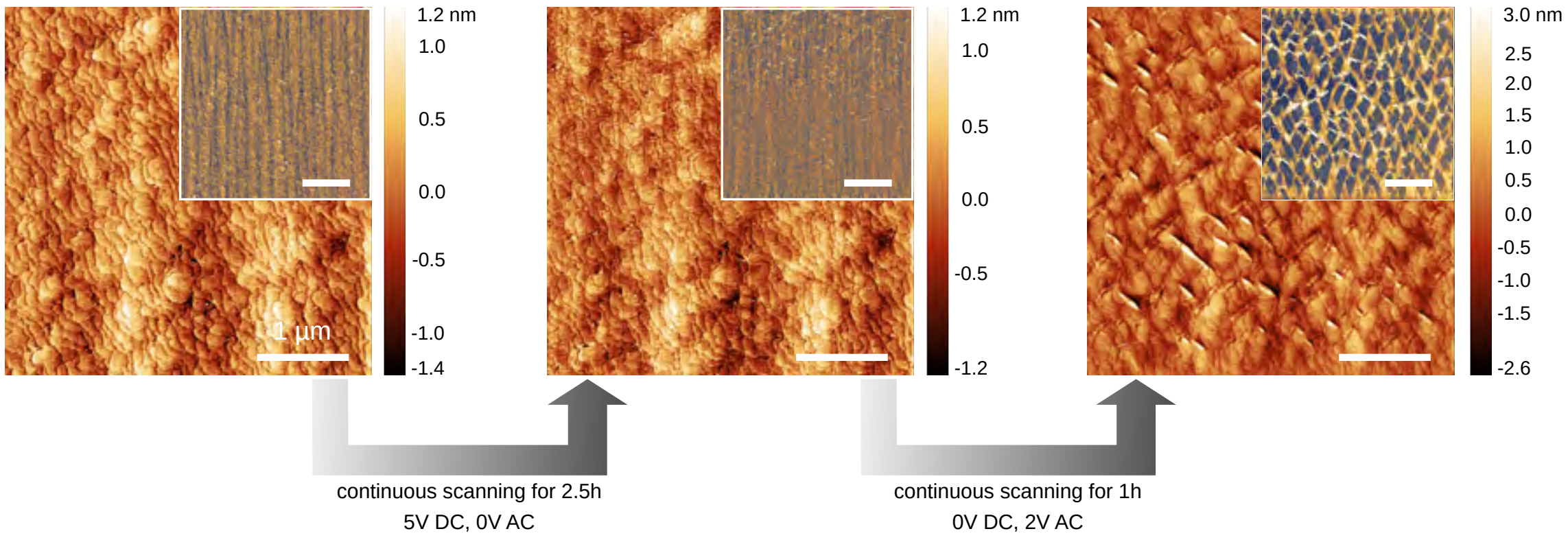


Figure S15: **Effect of DC bias on domain-wall expansion in $WO_3$.** Under continuous scanning with a DC bias of 5 V and no AC bias for 2.5 h, no changes in thin-film topography and the vertical PFM phase (inset) associated with the domain-wall-to-domain expansion can be discerned. When subsequently reapplying an AC bias of 2 V and no DC bias, the domain-wall expansion occurs again within 1 h and manifests both in the thin-film topography as diagonal trench-like features and the vertical PFM phase as diamond-shaped domains.

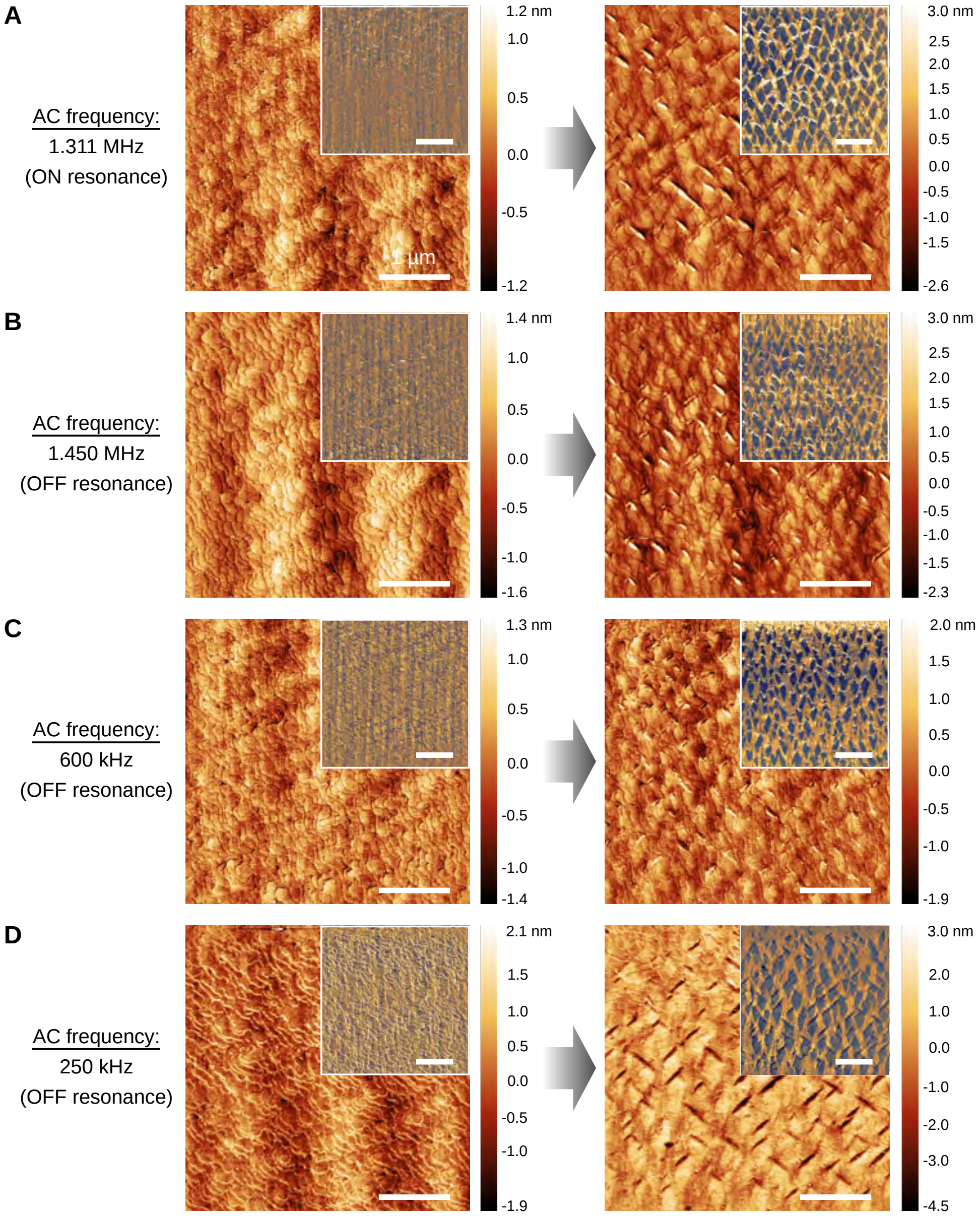


Figure S16: **Influence of the frequency of the AC bias on domain-wall expansion in $WO_3$.** (**A-D**) Topography and vertical PFM phase (insets) before (left column) and after applying an AC bias of $\geq 2$ V (right column) of varying frequencies for $\sim$40 min. We find that the observed phase transition and associated domain-wall expansion is independent of the AC bias frequency and whether it is chosen on the contact resonance (A) or at various values off-resonance (B-D). In all cases, we observe consistent changes in the thin-film topography and the vertical PFM phase upon continuous scanning. The phase transition is observed down to AC frequencies of 10 kHz (not shown).

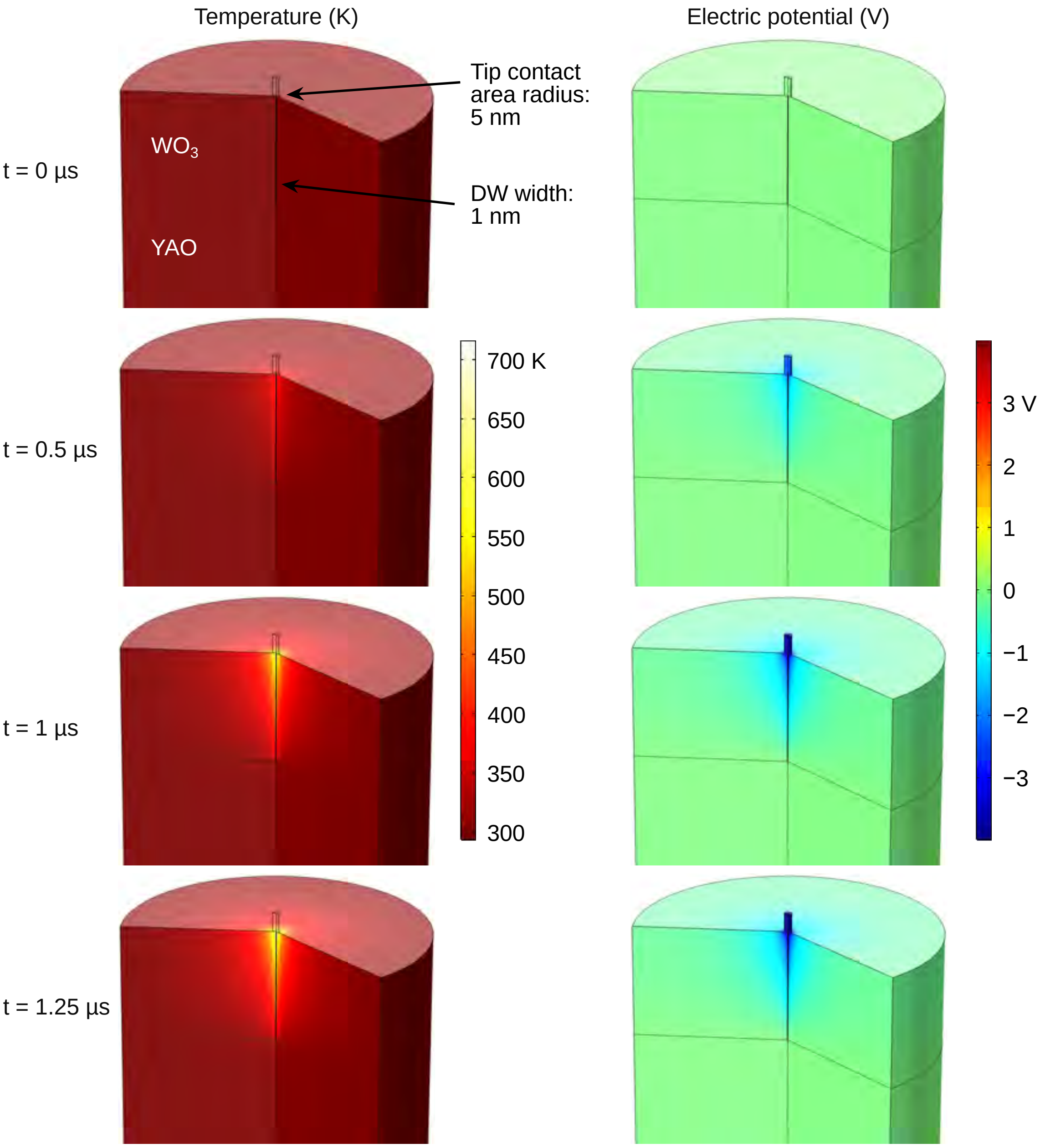


Figure S17: **Finite-element modeling of temperature and voltage profiles in $WO_3$ films.** The calculations assume a radius of the tip-sample contact area of 5 nm comparable with previous studies [18], a domain-wall width of 1 nm [13], an AC voltage amplitude of 4 V, and an AC electric field frequency of 400 kHz. Similar to the experimental setup, the calculation assumes the voltage being applied between the tip and a grounded silver spot on the surface of the $WO_3$ film, separated from the tip by several mm. The two columns show the resulting temperature (left) and voltage (right) evolution at different time stamps. After a time of 1.25 µs (well within the duration the tip resides at each spot of ∼1 ms), the peak electric potential of 4 V is reached under the PFM tip and results in a maximum temperature of about 700 K at the location of the domain wall.

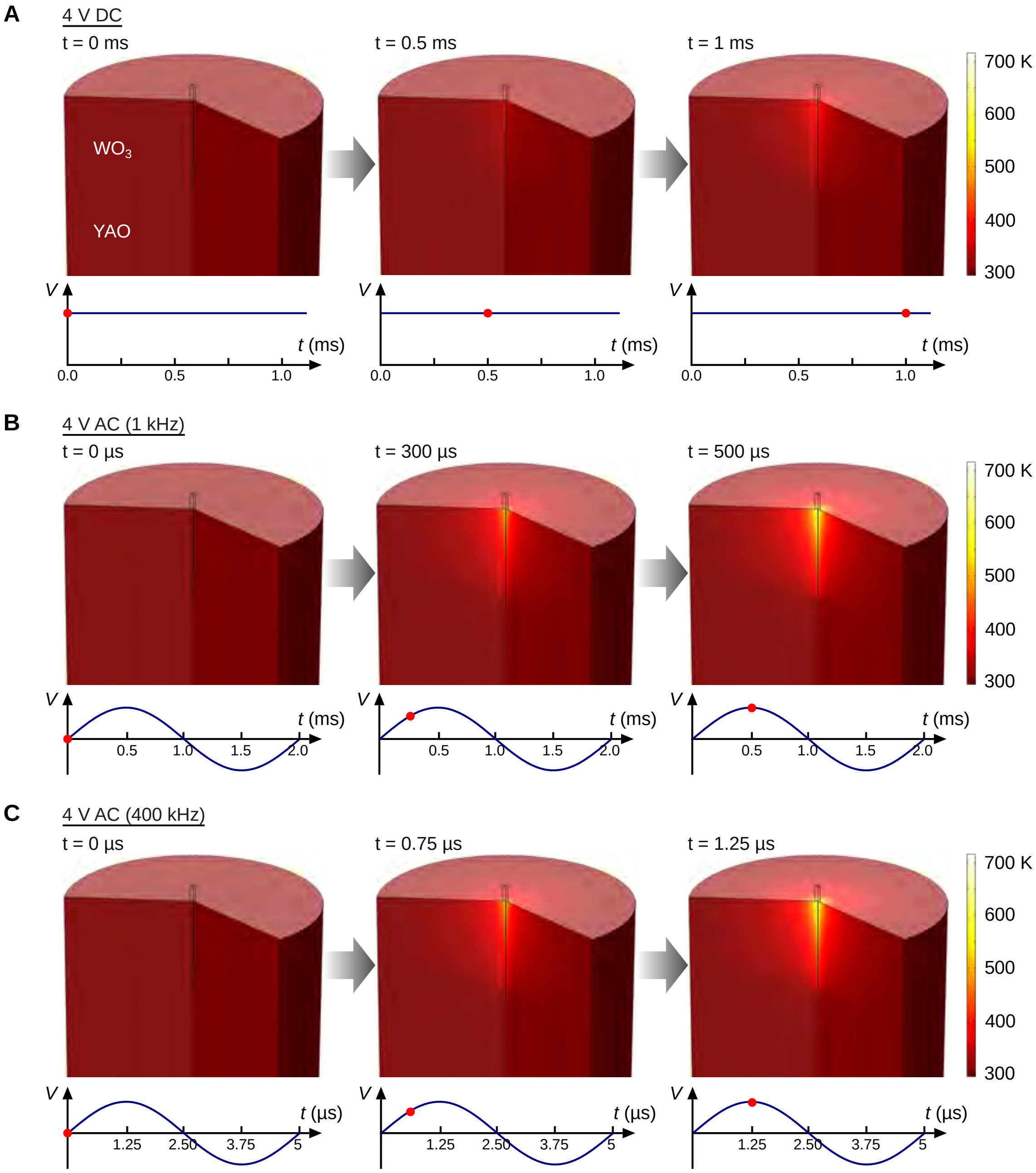


Figure S18: **Comparison of tip-applied AC and DC biases on the temperature profile across the $WO_3$ films.** (**A-C**) Finite-element modeling of the temperature profiles at different time stamps upon application of (A) a 4 V DC bias, (B) a 4 V AC bias at 1 kHz, and (C) a 4 V AC bias at 400 kHz to a PFM tip located on top of a conducting $WO_3$ domain wall. Within the duration of ∼1 ms the PFM tip resides on a single domain wall during scanning, a maximum temperature of around 700 K is reached for both AC field frequencies (B,C). In contrast, for an applied DC bias of 4 V a maximum temperature of 450 K is obtained.

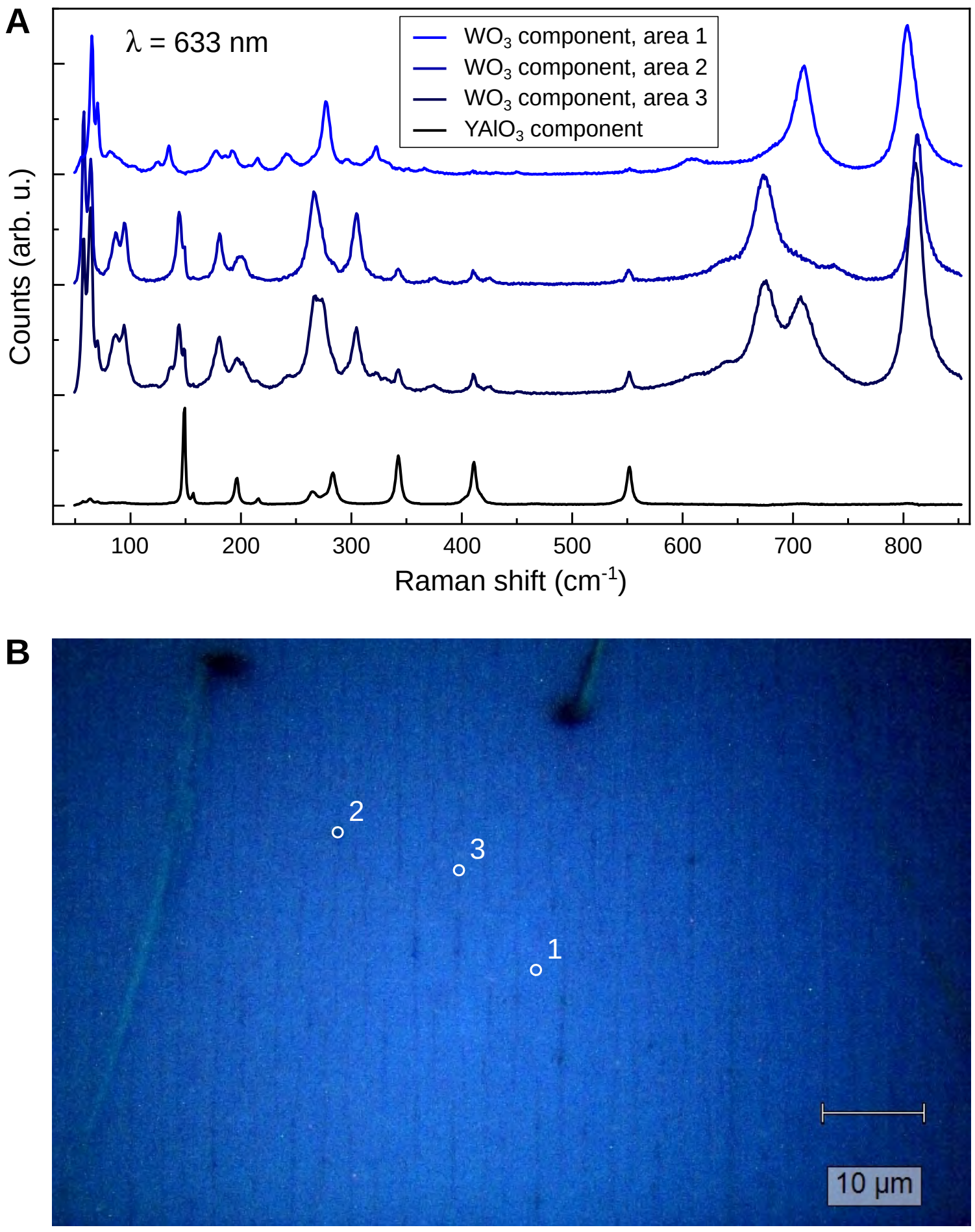


Figure S19: **Room-temperature Raman spectra on $WO_3$ after prior heating to** 675 °C**.** (**A**) NMF $WO_3$ spectral components originating from different regions of the $WO_3$ film, as marked in (B), measured with a laser excitation wavelength of 633 nm. The $YAlO_3$ substrate NMF component is shown as a reference. Depending on the sample location, clear spectral differences can be distinguished. Whereas at location 1, the characteristic spectrum of the pristine triclinic film is evident, location 2 features almost exclusively the characteristic peak of the tetragonal poled phase at 672 $cm^{-1}$. Location 3 contains a mixture of spectral contributions from both the pristine triclinic and the poled tetragonal phase. (**B**) Optical microscope image acquired with a x100 objective for a representative $WO_3$ sample area after prior sample heating to 675 °C. The broadened stripe-domain walls appear as vertical dark stripes in the image. The appearance of needle-like crystallites at the top half of the image indicates the irreversible sample degradation after heating.

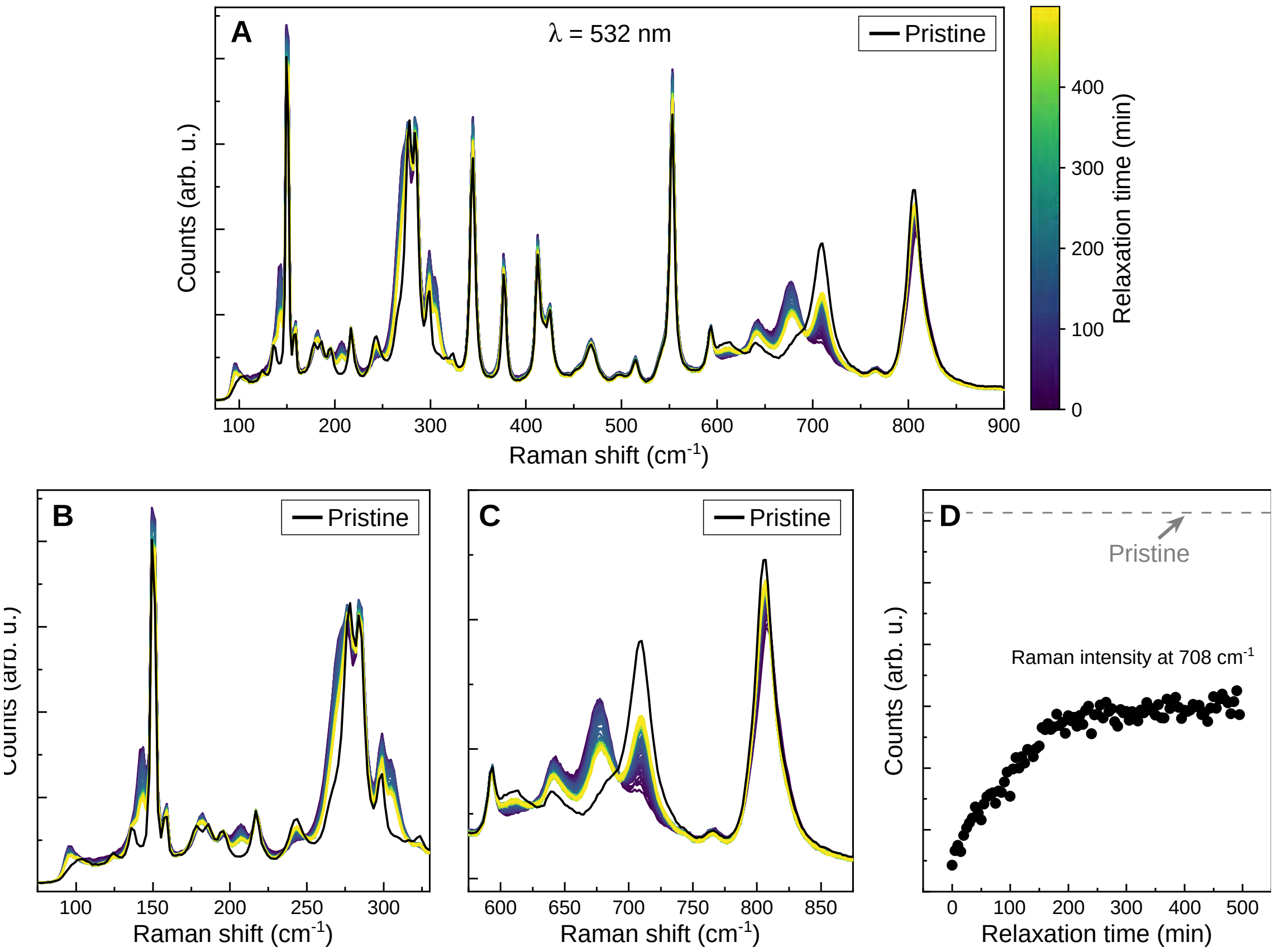


Figure S20: **Relaxation of the Raman spectra after domain-wall expansion in $\mathbf{WO_3}$.** (**A**) Evolution of the Raman spectra on a poled $WO_3$ area over time. The spectra correspond to the raw data acquired with an excitation wavelength of 532 nm without NMF into substrate and film components. For reference, the Raman spectrum of the pristine triclinic phase is shown in black. Note that a relaxation time of 0 min does not correspond to the end of PFM poling, but instead is approximately 5 h after PFM poling. (**B,C**) Enlarged views of the (B) low-frequency and (C) high-frequency spectral regions during relaxation. (**D**) Extracted scattered Raman intensity at 708 $cm^{-1}$ as a function of the relaxation time. Note that the Raman intensity reaches a quasi-equilibrium state without fully recovering the peak intensity of the pristine film. This is consistent with the remanent change of the Raman spectra after heating to 675 °C (Fig. S19).

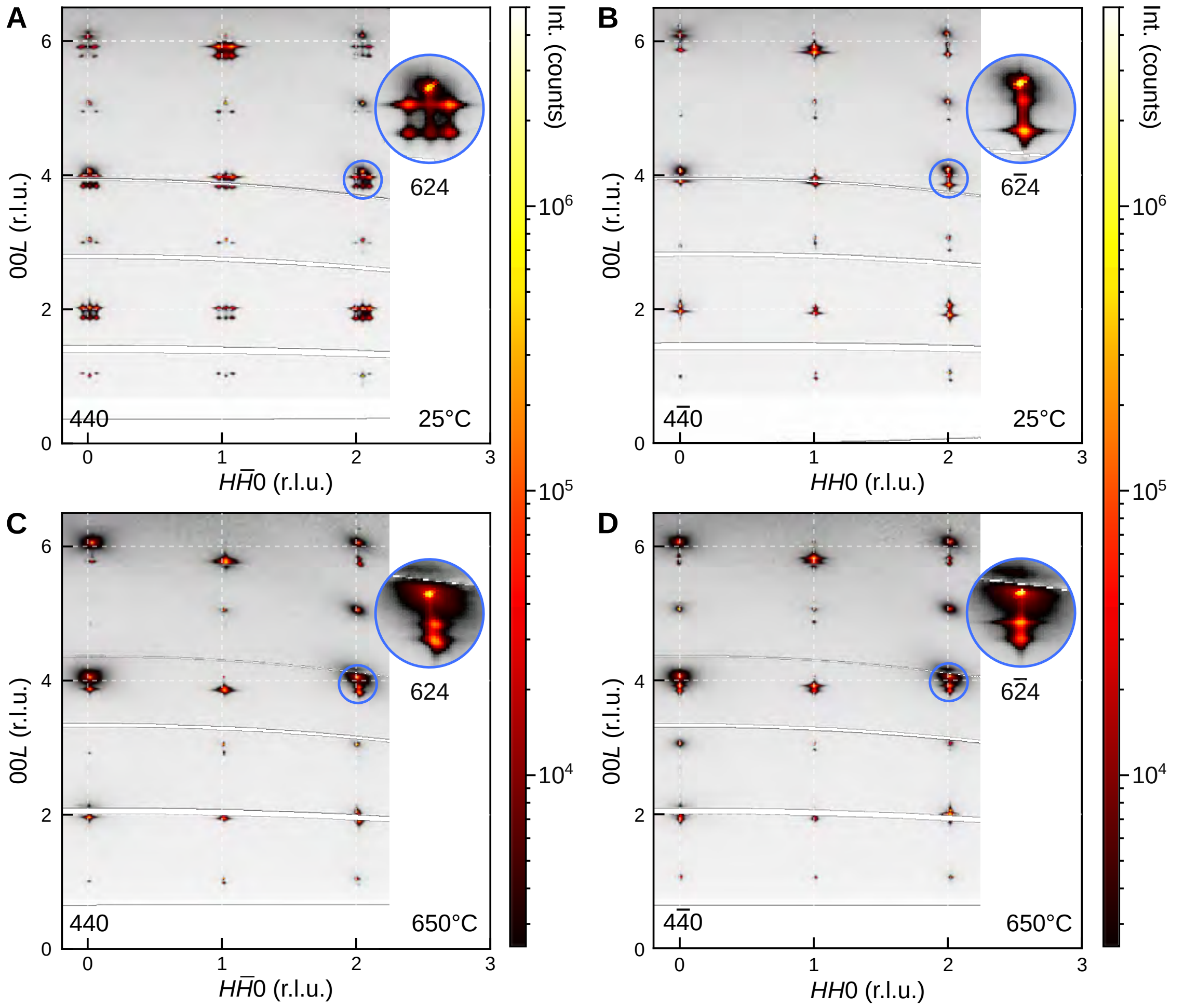


Figure S21: **Synchrotron-based large-area 3D reciprocal space surveys to elucidate the high-temperature phase of $WO_3$.** (**A,B**) Out-of-plane (A) $HH0$ with origin 440 and (B) $H\overline{H}0$ with origin $4\bar{4}0$ reciprocal space reconstructions at room temperature. In the enlarged view of (A) around the 624 peak, we observe the four-fold peak splitting as a result of the four triclinic structural variants present at room temperature. [13] In the perpendicular direction in (B), the projections of two triclinic domains coincide, giving rise to a two-fold peak splitting of the $6\bar{2}4$ peak. (**C,D**) Equivalent out-of-plane reconstructions as in (A,B) at a temperature of 650 °C. Here, the transition from triclinic to tetragonal results in equivalent diffraction patterns, as highlighted in the $6\bar{2}4$ reflections in the enlarged views.



\end{document}